\documentclass[a4paper,amsmath,amssymb,floatfix,pra,footinbib,twocolumn,showpacs,preprintnumbers]{revtex4}

\usepackage[pdftex]{graphicx}
\usepackage{dcolumn}
\usepackage{bm}
\usepackage{dsfont}
\usepackage{color}

\usepackage{url}
\usepackage[colorlinks=true ,citecolor=blue]{hyperref}
\usepackage{amsmath,amssymb}

\newcommand{\ttt}{$\mathcal{T}_3$~}

\newcommand{\ii}{\text{i}}
\newcommand{\ee}{\text{e}}

\newcommand{\SO}{\text{SO}}

\newcommand{\yy}{{\xi}}
\newcommand{\eps}{\varepsilon}

\newcommand{\ket}[1]{|#1\rangle}

\newcommand{\comment}[1]{}
\newcommand{\bcom}[1]{\textcolor{blue}{\textsf{\textbf{#1}}}}

\begin{document}



\title{Topological Phases for Fermionic Cold Atoms on the Lieb Lattice}
\author{N. Goldman}
\email{ngoldman @ ulb.ac.be}
\affiliation{Center for Nonlinear Phenomena and Complex Systems - Universit\'e Libre de Bruxelles (U.L.B.), Code Postal 231, Campus Plaine, B-1050 Brussels, Belgium}
\author{D.~F. Urban}
\affiliation{Physikalisches Institut, Albert-Ludwigs-Universit\"at, D-79104 Freiburg, Germany}
\author{D. Bercioux}
\affiliation{Freiburg Institute for Advanced Studies, Albert-Ludwigs-Universit\"at, D-79104 Freiburg, Germany}
\affiliation{Physikalisches Institut, Albert-Ludwigs-Universit\"at, D-79104 Freiburg, Germany}

\date{\today}

\begin{abstract}

We investigate the properties of the Lieb lattice, i.e a face-centered square lattice, subjected to external gauge fields. We show that an Abelian gauge field leads to a peculiar quantum Hall effect, which is a consequence of the single Dirac cone and the flat band characterizing the energy spectrum. Then we explore the effects of an intrinsic spin-orbit term -- a non-Abelian gauge field -- and  demonstrate the occurrence of the quantum spin Hall effect in this model. Besides, we obtain the relativistic Hamiltonian describing the Lieb lattice at low energy and derive the Landau levels in the presence of external Abelian and non-Abelian gauge fields. Finally, we describe concrete schemes for realizing these gauge fields with cold fermionic atoms trapped in an optical Lieb lattice. In particular, we provide a very efficient method to reproduce the intrinsic (Kane-Mele) spin-orbit term with assisted-tunneling schemes. Consequently, our model could be implemented in order to produce a variety of topological states with cold-atoms.

\end{abstract}

\maketitle

%
%
\section{Introduction}

During the last decades, topology has influenced many fields of physics through the renewed description of various phenomena. In condensed-matter physics, topological invariants | known as Chern numbers | have played an important role in the description of the integer quantum Hall effect (IQHE)~\cite{Klitzing1986}. Here,  the quantized Hall conductivity of a two-dimensional (2D) electron system is expressed as a sum $\sigma_\text{H}=R_{\text{K}}^{-1} \sum_{E_n < E_\text{F}} N_{\text{ch}} (E_n)$
of Chern numbers $N_{\text{ch}} (E_n)$ that are integers associated with the energy band $E_n$~\cite{Thouless1982,Kohmoto1985}.
Here $R_{\text{K}}$ is von Klitzing's constant and $E_\text{F}$ denotes the Fermi energy assumed to lie inside a gap of the bulk energy spectrum.  Furthermore, it has been proven that the sum of Chern numbers is expressing the number of gapless \emph{edge states} located inside the bulk energy gaps. These edge states carry the current in the IQHE~\cite{Hatsugai1993,Qi2006}.

The breaking of time-reversal symmetry (TRS)  due to external magnetic fields plays a crucial role for the topological interpretation of the IQHE~\cite{Haldane1988}.  Recently, the discovery of the so-called quantum spin-Hall effect (QSHE) has lightened a new path for the investigation of systems where TRS is preserved~\cite{Kane2005,Kane2005bis,Kane2006,Bernevig2006,Bernevig2006bis,Qi2008}. The QSHE manifests itself in insulating systems that show a non-trivial Z$_2$ index~\cite{Kane2005}.
These so-called \emph{topological insulators} are characterized by the presence of \emph{spin-filtered} edge states in the gaps of the bulk energy spectrum. Because of TRS invariance, the spin-up and spin-down states move in opposite directions along the edge of the system. As a consequence, the total charge current as well as the associated Chern numbers are zero~\cite{Avron1988,Qi2008}. Yet, a spin-Chern number has been introduced in order to measure the spin-transport~\cite{Sheng2006} and to distinguish the Z$_2$ class of the system~\cite{Fukui2007}.

The interplay between the lattice topology and the QSHE has been the focus of various recent investigations. In particular, the QSHE has been studied  for the Kagome~\cite{Guo2009,Liu2010}, the Lieb and Perovskite~\cite{Franz:2010}, the honeycomb~\cite{Kane2005,Ruegg2010}, the square~\cite{Stanescu2010,goldman:2010}, the $\mathcal{T}_3$~\cite{bercioux:2010}, the checkerboard~\cite{Sun2009}, the pyrochlore~ \cite{Guo2009PRL}, diamond~\cite{Fu2007} and the square-octogon~\cite{Kargarian2010} lattices.

In this context, some lattice models are of particular interest as they show dispersionless energy bands. These \emph{flat bands} correspond to a macroscopic number of degenerate localized states.
Originally, flat bands played a fundamental role in magnetism, as they were shown to accompany the occurance of ferromagnetic ground states in multi-band Hubbard models \cite{Tasaki2008,Lieb1989,Mielke1991}. More recently, the existence and the robustness of these special bands have been extensively studied in a vast family of frustrated hopping models \cite{Bergman2008,Green2010} and for the case of electron localization due to magnetic fields and spin-orbit interactions~\cite{localization}.  Interestingly, singular touchings between flat and dispersive bands have been shown to be topologically protected by real-space loops \cite{Bergman2008}.  On the face-centered square lattice | also known as  Lieb lattice | a flat band touches two linearly dispersing bands, \emph{i.e.}, the flat band intersects a single Dirac point, and the low-energy regime is described by a quasi-relativistic equation for spin-1 fermions~\cite{schen:2010,Dagotto}.

Nowadays, various lattices can be engineered using cold atoms trapped by electromagnetic fields \cite{Lewenstein2007,Bloch2008,schen:2010,apaja:2010,grynberg:1993,bercioux:2009}. In particular, the realization of topological states of matter with cold fermionic atoms appears to be a realistic and attractive goal from the experimental point of view \cite{Stanescu2010,goldman:2010,Liu:2010}. A significant advantage of these experiments is the full control of a wide range of system parameters as, \emph{e.g.}, lattice geometry, interaction and disorder. In these experiments, engineered gauge fields allow to mimic the effects of magnetic fields \cite{Lin2009,Spielman2009,Jaksch2003} or spin-orbit interactions (SOIs) \cite{Stanescu2010,goldman:2010,Stanescu2008,Spielman2010, Wang2010b,Ho2010,Juze2010}. These gauge fields can be generated by spatially-varying laser or magnetic fields which modify particle-hopping via non-trivial Berry's phases \cite{Juze2005,Dalibard2010}. Recent experiments have implemented light-induced  external gauge fields and reproduced the physics of charges subjected to electric or magnetic fields \cite{Lin2009,Spielman2009,Lin2010}.  Moreover, with such a setup one expects to observe several fundamental phenomena including the Hofstadter butterfly \cite{Jaksch2003,Hofstadter1976}, atomic analogues of the quantum Hall effects \cite{Goldman2009,Sorensen2004},  relativistic physics \cite{Merkl2008,Goldman2009bis}, and vortex structures \cite{Gunter2009,Lin2009,Lim2008}.   Optical-lattice setups also allow to consider a generalization of the ongoing experiments, namely the implementation of non-Abelian gauge fields \cite{Osterloh2005, Ruseckas2005,Gerbier2009,Spielman2010}. In particular, non-Abelian gauge fields acting on multi-level atomic systems could mimic SOI \cite{Stanescu2008,Spielman2010, Wang2010,Ho2010,Juze2010}, paving the way to study the spin Hall \cite{Zhu2006} and quantum spin Hall effects \cite{goldman:2010,Liu2010}. Very recently, a concrete proposal of an optical Lieb lattice for cold atoms has been presented \cite{apaja:2010}. In the later work, Apaja \emph{et al.} have shown that a fermionic cloud expanding after the release of the harmonic trap should show clear signatures of the flat band's localized states. Finally, the existence of flat bands with non-trivial topological order has been demonstrated \cite{Tang2010}, contradicting the belief that non-dispersive bands were associated to vanishing Chern numbers \cite{Green2010}. \\

Motivated by the possibility to engineer an optical Lieb lattice for cold fermionic atoms, we investigate the emergence of topological properties for various configurations of synthesized gauge fields. We first provide an original analysis of a peculiar IQHE, in the case where a uniform magnetic field is present in the Lieb lattice. We then explore the effects of an intrinsic spin-orbit term \cite{Kane2005} and show how it leads to quantum spin Hall states. In this framework, we extend the seminal work of Ref.~[\onlinecite{Franz:2010}] and derive an effective Hamiltonian describing the low-energy regime. This Weyl-like Hamiltonian leads to a three-component quantum equation that resembles the relativistic equation for spin-1 particles. Besides, we obtain the Landau levels in the presence of an external magnetic field and spin-orbit interaction. Finally, we discuss the optical-lattice realization of this Lieb system and propose realistic methods for creating Abelian (magnetic) and non-Abelian (spin-orbit) gauge fields. We show that the Lieb lattice is particularly suited to reproduce the intrinsic spin-orbit term introduced by Kane and Mele \cite{Kane2005}. The later, which involves complex spin-dependent \emph{next}-nearest-neighbour hoppings, can be simply decomposed into nearest-neighbour hopping on a square sublattice. This elegant idea is a non-Abelian generalization of the method proposed in Ref.~[\onlinecite{Liu2010}] for generating the Abelian Haldane-type gauge field.

%
%
\section{The Lieb lattice and topological phases in external fields}
%
%
\begin{figure}
	\centering
 	\includegraphics[width=0.7\columnwidth]{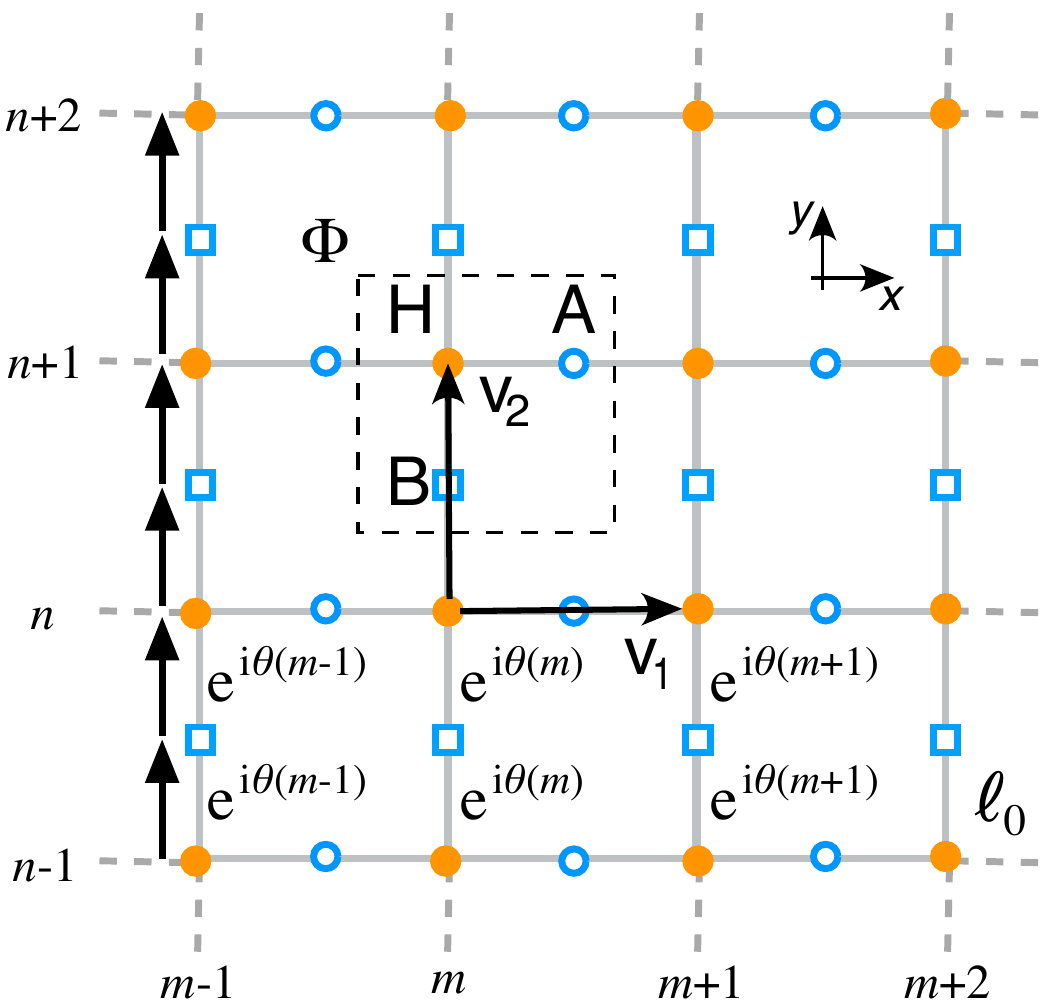}
	\caption{\label{fig:lattice} The face-centered square lattice or Lieb lattice. The Peierls phases $e^{i \theta(m)}$, where $\theta (m) = \pi \Phi m$, are associated to a uniform magnetic flux per plaquette $\Phi$ and are indicated by vectical black arrows. We set $x=2 m \ell_0$ and $y=2 n \ell_0$.}
\end{figure}
%
%

\begin{figure}
	\centering
 	\includegraphics[width=.8\columnwidth]{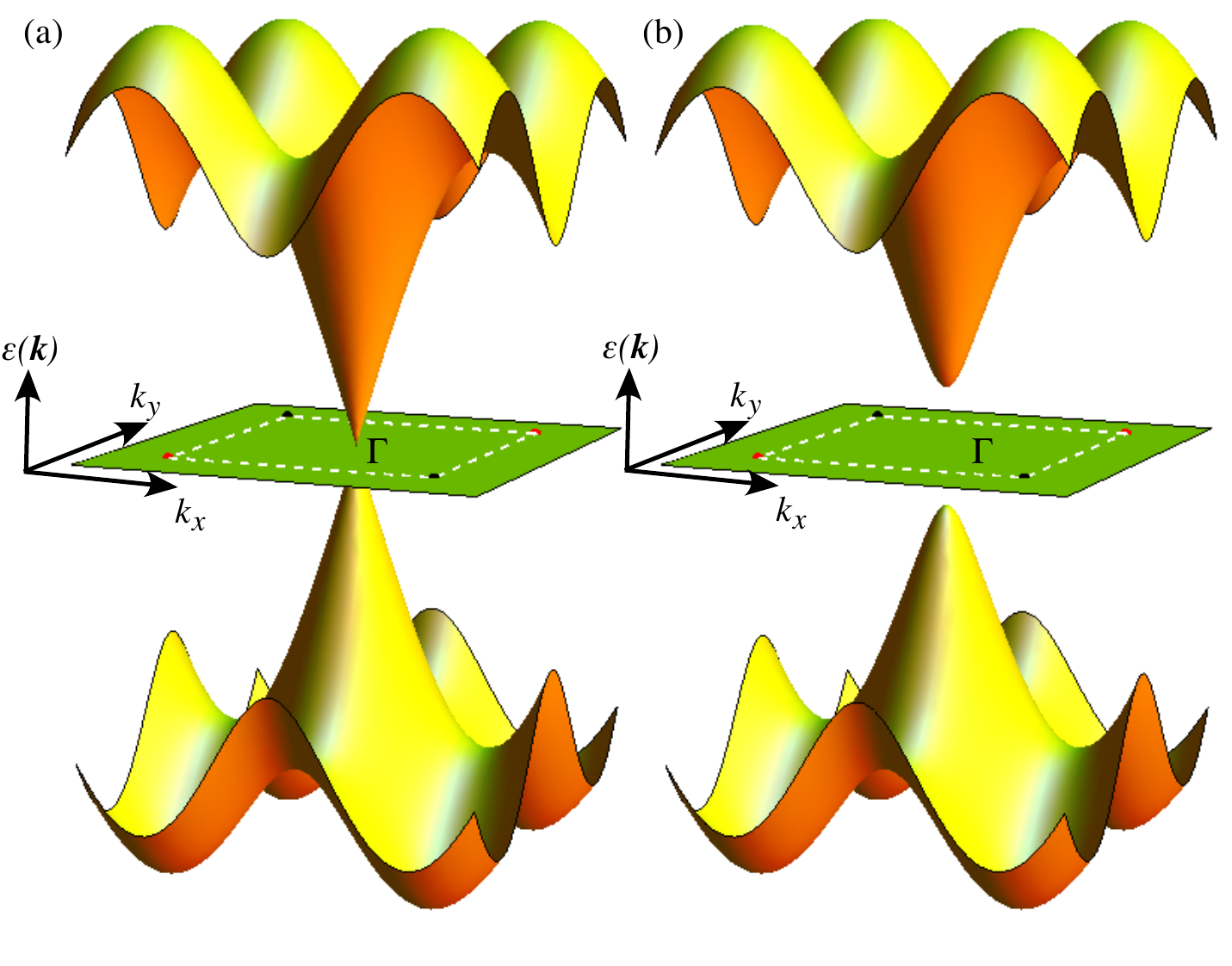}
	\caption{\label{fig:bulk:spectrum} Energy spectrum of the Lieb lattice as a function of the momentum $\bm{k}=(k_x,k_y)$. The dashed lines delimit the first Brillouin zone. Panel (a): spectrum without external fields. Panel (b): energy spectrum in the presence of spin-orbit interaction $t_\text{SO}=0.1\ t$.}
\end{figure}

We consider the face-centered square (Lieb) lattice, which is shown in Fig. \ref{fig:lattice}. This lattice has a unit cell characterized by three lattice sites, hereafter referred to as H, A and B. Site H has four nearest-neighbors (NN), namely two A and two B sites. On the contrary the A and B sites have only two NN H sites.
The bulk properties of the Lieb lattice can be analyzed within a tight-binding (TB) approximation. In this limit, the Hamiltonian of the system can be written as $\mathcal{H}_0=t \sum_{\langle i,j \rangle\alpha} c^\dag_{i \alpha} c_{j\alpha}$ with spin independent NN hopping amplitude $t$. Here, $c_{j \alpha}^\dag(c_{j \alpha})$ is the  creation (annihilation) operator for a particle with spin direction $\alpha$ on the lattice site $j$. In absence of external fields the problem can be diagonalized exactly and the spectrum reads
%
%
\begin{subequations}\label{eq:spectrum}
\begin{align}
\varepsilon_0 (\bm{k}) & = 0,\label{spec:zero}\\
\varepsilon_\pm (\bm{k})  & = \pm t \sqrt{4 + 2 \cos(\bm{v}_1\cdot \bm{k}) + 2 \cos(\bm{v}_2\cdot \bm{k})},
\end{align}
\end{subequations}
%
%
where $\bm{k}=(k_x,k_y)$ and $\bm{v}_{1/2}$ are the lattice vectors, c.f. Fig.~\ref{fig:lattice}.
The bulk energy spectrum is shown in Fig.~\ref{fig:bulk:spectrum}a | it depicts  two identical, electron-hole symmetric  branches $\varepsilon_\pm$. Moreover, the Lieb lattice presents a unique non-dispersive band  at the charge neutrality point (CNP). This band is rooted in the lattice topology, which allows for insulating states with finite wave function amplitudes on the A and B sites and vanishing amplitudes on H sites. This property holds also when hopping to higher-order neighbors is allowed.  Note that the three bands touch at the center of the first Brillouin zone, which we set for simplicity at $\Gamma=\pi/2\ell_0(1,1)$. The resulting properties of carriers in proximity of the $\Gamma$ point are investigated in Sec.~\ref{LWA}.
%
%
%
%

%
%
\subsection{Uniform magnetic field and quantum Hall phases}

We now study the effects of a uniform magnetic field $\bm{B}= B \hat{z}$ on the spectral and transport properties of the Lieb lattice. We consider the Landau gauge
%
%
\begin{equation}\label{gaugeu1}
\bm{A}= \left( 0, B x , 0 \right) = \left( 0, \frac{\pi \Phi m}{\ell_0} , 0 \right) ,
\end{equation}
%
%
where $\Phi= \Phi_0^{-1} \int_{\square} \bm{B} \cdot \text{d}\bm{S}$ is the number of magnetic flux quanta $\Phi_0$ per plaquette and $x=2 m \ell_0$. Hereafter we use the notation $(m,n, \zeta)$, with $\zeta=\{\text{A,B,H}\}$, to label the lattice sites.
The gauge field $\bm{A}$ modifies the hopping along the $y$ direction through $x$-dependent Peierls phases $t \to t \ee^{\ii \theta(m)}$, where the phase reads
%
%
\begin{equation}\label{peierls}
\theta (m) = \int_{(m, n,\text{H})}^{(m, n+1,\text{B})} \bm{A} \cdot \text{d}\bm{l}= \int_{(m, n,\text{H})}^{(m, n,\text{B})} \bm{A} \cdot \text{d}\bm{l}= \pi \Phi m ,
\end{equation}
%
%
as illustrated in Fig.~\ref{fig:lattice}. Here, the integrations are performed along the links connecting the neighboring B and H  sites.

\begin{figure}[tbp]
\begin{center}
\includegraphics[width=0.9\columnwidth]{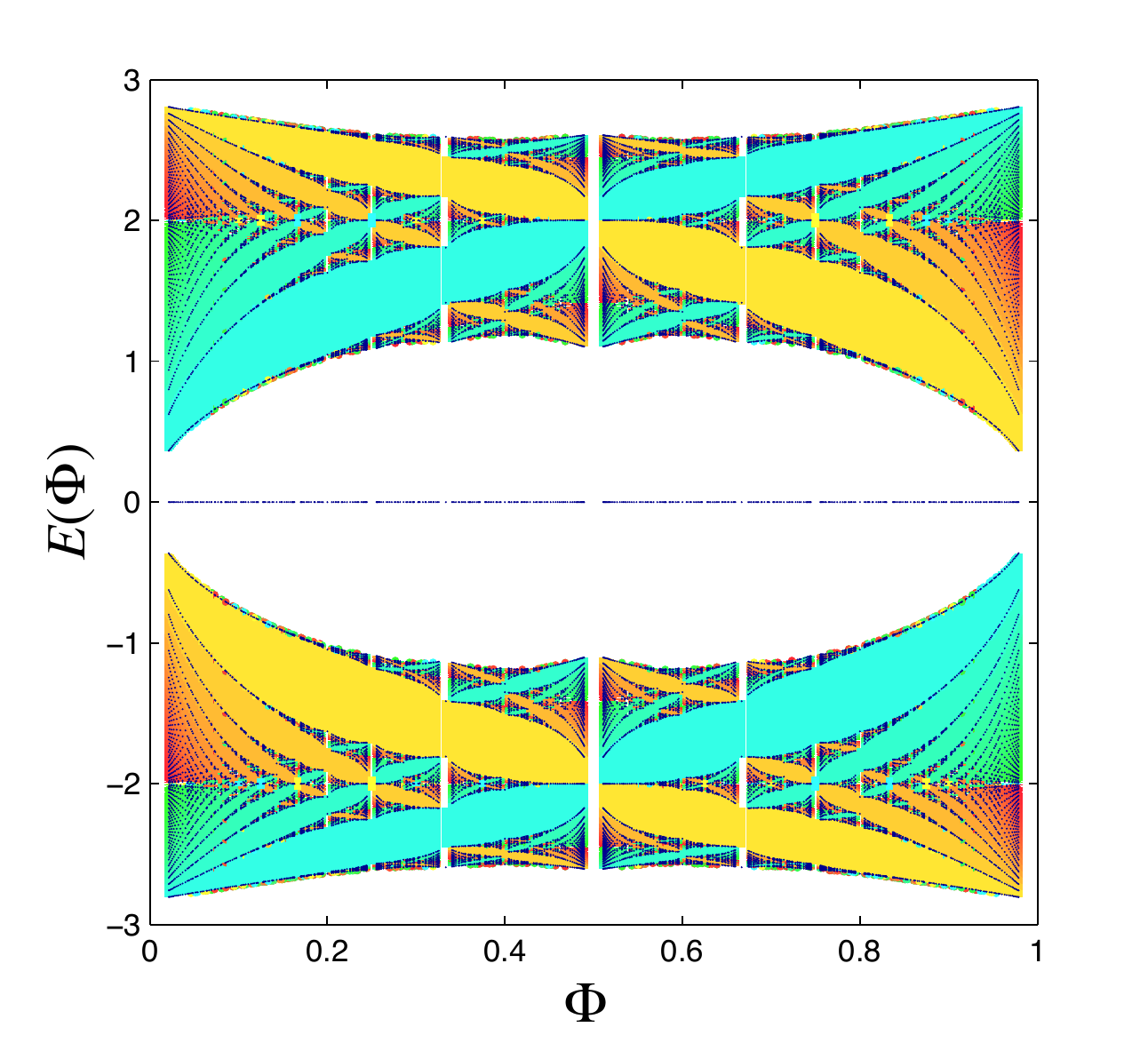}
\end{center}
\caption{\label{fig:butterfly} Spectrum $E=E(\Phi)$ and phase diagram for $\Phi ={p}/{q}$ with $q<47$. The eigenvalues are dark blue dots forming two successive butterflies. Gaps are filled with cold [resp. warm] colors according to the related positive [resp. negative] values of the quantized conductivity $\sigma_{xy}$. The white gaps located around $E=0$ correspond to $\sigma_{xy}=0$.}
\end{figure}

Setting $\Phi=p/q$, where $p$ and $q$ are mutually prime integers, the system becomes $q$-periodic along the $x$ direction. By considering periodic boundary conditions, it is possible to diagonalize the resulting  $3q \times 3q$ spectral problem. This leads to the fractal energy spectrum shown in Fig.~\ref{fig:butterfly}. As a function of the flux $\Phi$, the allowed energies depict two Hofstadter butterflies separated by a flat band at $E=0$ \cite{footnote1}. This specific band is reminiscent of the flat band obtained at zero magnetic field.

The fractal energy spectrum of lattices subjected to uniform magnetic fields are intimately related to the IQHE~\cite{kohmoto:1989,hasegawa:1990}. When the Fermi energy $E_{\text{F}}$ is located in a spectral gap, the Hall transverse conductivity of the system is quantized. This relation is supported by a Diophantine equation~\cite{kohmoto:1992} which expresses the quantized Hall conductivity in terms of the magnetic flux and the position of the gap: In the $r$--th gap, the Hall conductivity is given by $\sigma_{xy}=(e^2/h) t_r$, where the integers $(t_r , s_r)$ satisfy
%
%
\begin{equation}
r = p t_r + q s_r . \comment{, \qquad \Phi=p/q .}
\end{equation}
%
%
In general, the solutions $(t_r , s_r)$ are not unique and additional criteria are needed in order to find quantized values of the Hall conductivity~\cite{kohmoto:1992}.  The integer $t_r$ also has a topological interpretation, since it represents the sum of Chern numbers characterizing the bands below the Fermi energy $E_{\text{F}}$:
%
%
\begin{equation}
t_r = - \sum_{E_{\lambda} < E_{\text{F}}} N_\text{Ch} (E_{\lambda}) .
\end{equation}
%
%

In order to investigate the quantum Hall phases in the Lieb lattice, we have numerically evaluated the Chern numbers $N_\text{Ch} (E_{\lambda})$ using the method of Ref.~\cite{fukui:2005}.  We have verified that the integer $t_r$ satisfies the Diophantine equation with the specific condition $\vert t_r \vert \le q/2$~\cite{kohmoto:1992}. The full phase diagram describing the integer quantum Hall effect for the Lieb lattice is drawn in Fig.~\ref{fig:butterfly}. It represents the infinitely many quantum Hall phases, characterized by the quantized transport coefficient  $\sigma_{xy}=(e^2/h) t_r$, inside the spectral gaps. The different positive [resp. negative] values of the Hall conductivity are designated by cold [resp. warm] colors. \\

\begin{figure}[tbp]
\begin{center}
\includegraphics[width=0.9\columnwidth]{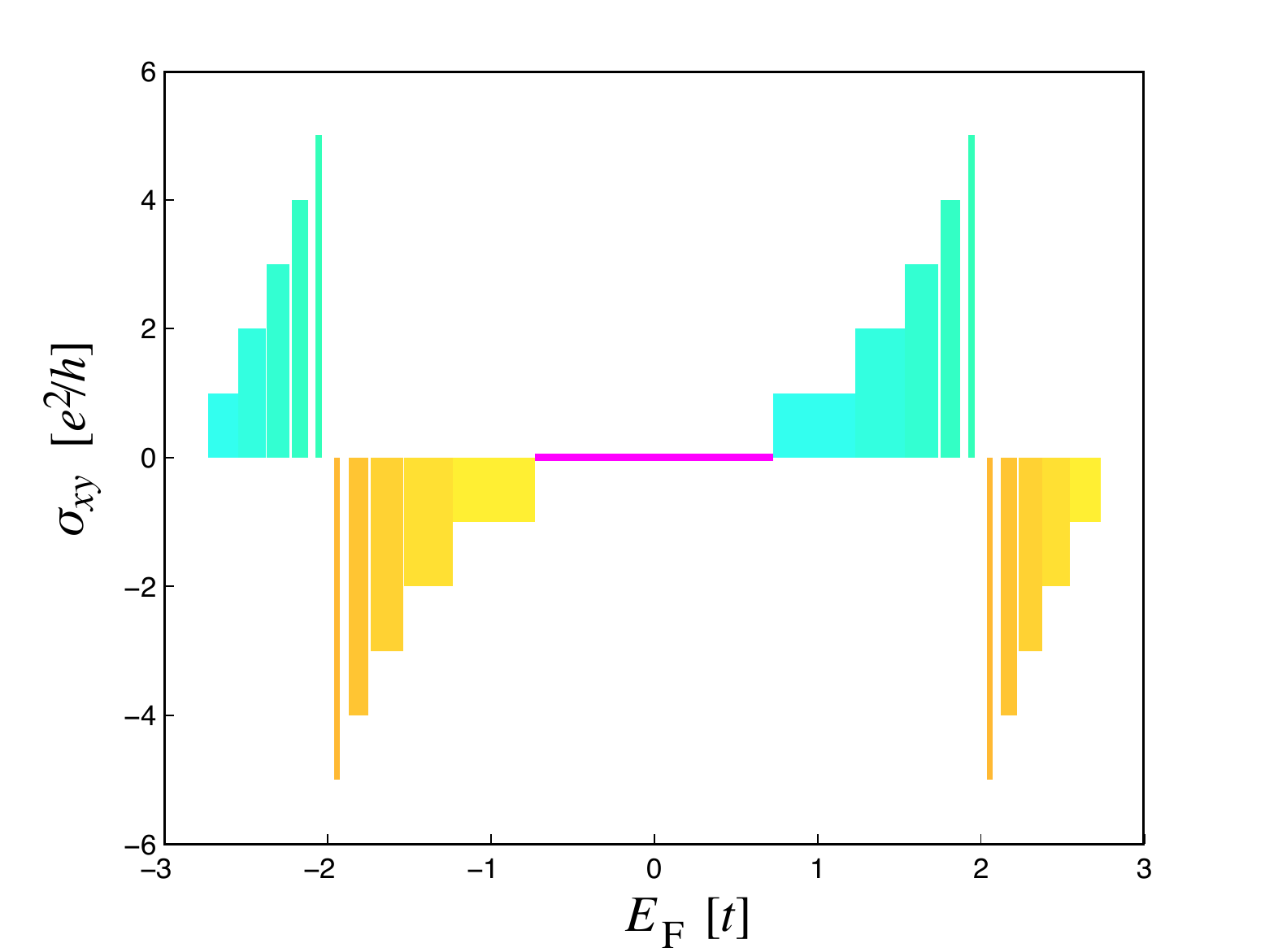}
\end{center}
\caption{\label{fig:Hall:cond} Hall conductivity $\sigma_{xy}(E_{\text{F}})$ as a function of the Fermi energy for $\Phi=1/11$. Cold [resp. warm] colors correspond to positive [resp. negative] values of the quantized conductivity, while the central plateau at $\sigma_{xy}=0$ is represented in magenta.}
\end{figure}

To identify the Hall plateaus stemming from the uniform magnetic field, we represent the Hall conductivity $\sigma_{xy}(E_{\text{F}})$ as a function of the Fermi energy in the low-flux regime $\Phi\ll1$ (cf. Fig.~\ref{fig:Hall:cond}).  In this regime, the quantized conductivity evolves monotonically but suddenly changes sign around the van Hove singularities (VHS) ~\cite{hatsugai:2006}, located at $E = \pm 2\ t$ (see the alternation of cold and warm colors in Figs.~\ref{fig:butterfly} and \ref{fig:Hall:cond}). Note that the gaps surrounding the topological flat band at $E=0$, correspond to normal band insulators with vanishing conductivity $\sigma_{xy}=0$, for all values of the flux $\Phi$. This is a consequence of the flat band's vanishing Chern number \cite{Green2010}. Most importantly, we observe that the Hall sequence presented in Fig. \ref{fig:Hall:cond} shows steps of $\Delta \sigma_{xy}=(e^2 /h)$. It is interesting to compare the latter result with the Hall sequences obtained for the $\mathcal{T}_3$ and honeycomb lattices \cite{bercioux:2010,hatsugai:2006}, which are characterized by steps of  $\Delta \sigma_{xy}=2 (e^2 /h)$ between the VHS. This major difference \cite{Lan2011} is due to the fact that the Lieb lattice is characterized by a \emph{single} Dirac-Weyl point, whereas the $\mathcal{T}_3$ and honeycomb lattices display \emph{two} Dirac cones.

%
%
%
%

%
%
%
%

%
%
\subsection{Spin-orbit interaction and quantum spin Hall phases}

In this Section, we study the effects of spin-orbit interactions (SOIs) on the properties of the Lieb lattice. Specifically, we introduce an intrinsic SOI term in the TB Hamiltonian, in analogy with the model of Kane and Mele for graphene~\cite{Haldane1988,Kane2005,Kane2005bis, min:2006,bercioux:2010,Franz:2010}. This term is modeled via a spin-dependent next-nearest-neighbor (NNN) hopping term
%
%
\begin{equation}\label{eq:SO}
\mathcal{H}_\text{SO} = \text{i}\,  t_\text{SO} \sum_{\alpha\beta}\!\!\sum_{\langle\langle k,l \rangle\rangle}
c_{k,\alpha}^\dag \left(\mathbf{d}_i\times\mathbf{d}_j\right)\cdot \bm{\sigma}_{\alpha\beta} \ c_{l,\beta}\, .
\end{equation}
%
%
The $\bm{\sigma}_{\alpha\beta}$ are matrix elements of the Pauli-matrices $\bm{\sigma}$ with respect to the final and initial spin states $\alpha$ and $\beta$ and $\mathbf{d}_{i/j}$ are the two displacement vectors of the NNN hopping process connecting sites $k$ and $l$.  Since in 2D lattices hopping is naturally restricted to in-plane processes, the SOI is effectively proportional to $\sigma_z$. Because of the unequal connectivity of A/B and H sites, the term~(\ref{eq:SO}) effectively induces hopping between  A and B sites only, \emph{i.e.}, other NNN-hopping processes cancel.
The spectrum is obtained by exact diagonalization~\cite{Franz:2010} and reads
%
%
\begin{subequations}
\begin{align}\label{spectrum:SOI}
\varepsilon^\text{(SO)}_0 (\bm{k})& = 0 \\
\varepsilon^\text{(SO)}_\pm(\bm{k}) & = 2 \left[ t^2 \left(\cos^2(\bm{v}_1\cdot\bm{k})+\cos^2(\bm{v}_2\cdot\bm{k})\right) \right. \\
& ~~~\left. +4 t_\text{SO} ^2 \sin^2(\bm{v}_1 \cdot \bm{k}) \sin^2(\bm{v}_2\cdot\bm{k})\right]^{\frac{1}{2}}\,.
\end{align}
\end{subequations}
%
%
The bulk energy spectrum is shown in Fig.~\ref{fig:bulk:spectrum}b for $t_\text{SO}=0.1\,t$. Due to its topological origin, the non-dispersive band at $E=0$ is not affected by SOI. However, this term opens two bulk energy gaps $\Delta_\text{gap}=4\  t_\text{SO}$ between the non-dispersive and the electron/hole branches, respectively.

The SOI has dramatic consequences on the transport properties of the Lieb lattice: as shown by Weeks and Franz, the gap $\Delta_\text{gap}$ allows for a topological insulating phase~\cite{Franz:2010}. The latter is characterized by a robust spin transport along the edges of the system. This quantum spin Hall phase, induced by the Haldane-type term~(\ref{eq:SO}), can thus be directly visualized when studying a finite piece of the lattice, \emph{i.e.}, by considering its edges.

\begin{figure}[tbp]
\begin{center}
\includegraphics[width=\columnwidth]{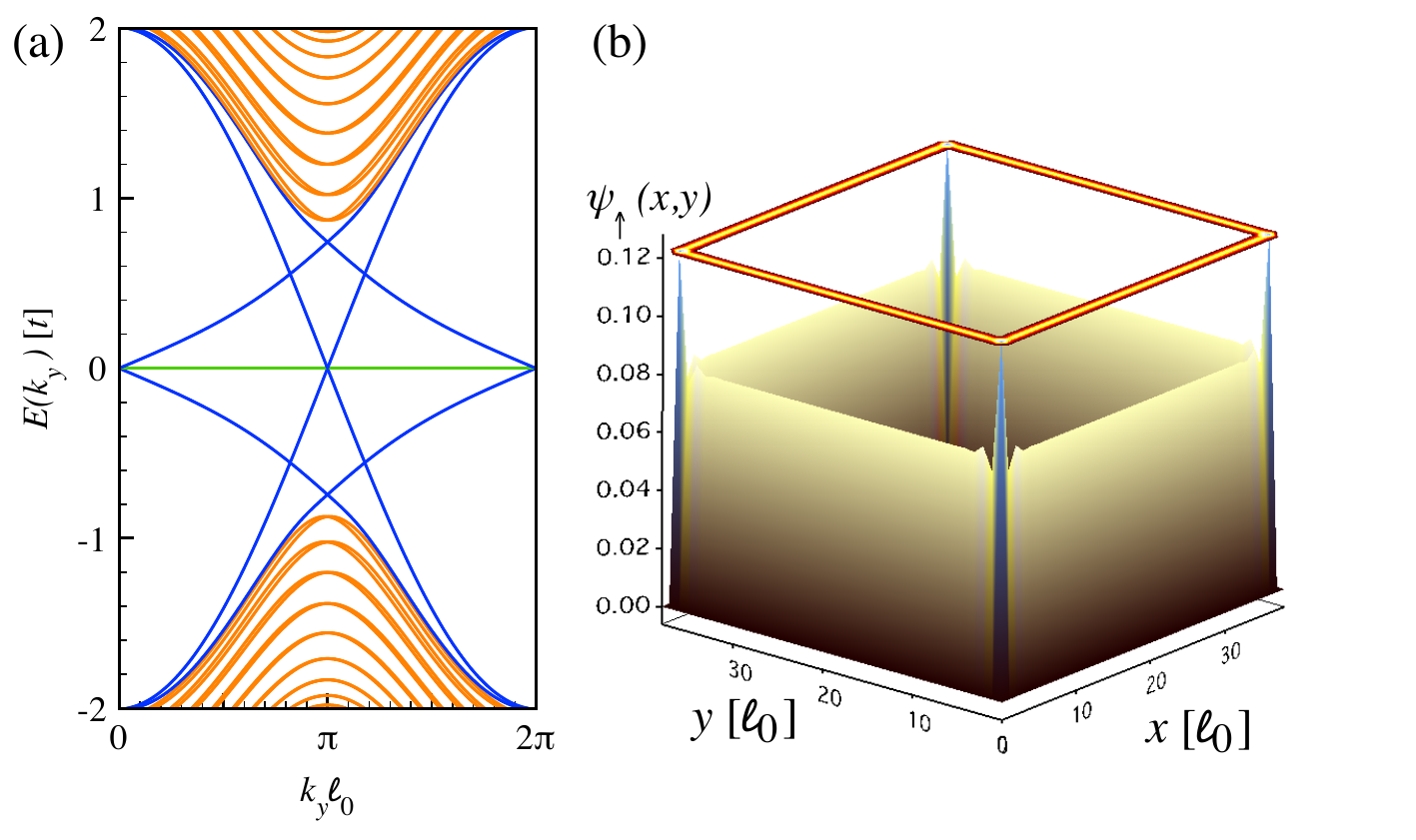}
\put(-147,110){$\vert$}
\put(-122,110){$\vert$}
\caption{\label{fig:spectrum:cylinder} Panel (a): Energy spectrum $E=E(k_y)$ in the cylinder geometry with $t_\text{SO}=0.2 t$. The bulk gap is traversed by gapless edge-states [i.e. helical states]. Panel (b): Edge-states amplitude $\vert \psi_{\uparrow} (x,y) \vert ^2$ for the open Lieb lattice with $39 \times 39$ sites, straight edges and $t_\text{SO}=0.5 t$. This localized eigenstate satisfies $H \psi_{\lambda} (x,y)= E_{\lambda} \psi_\lambda (x,y)$ and corresponds to the energy $E_\lambda=0.5 t$ which lies within the gap.}
\end{center}
\end{figure}

A standard method consists in diagonalizing the TB Hamiltonian with periodic boundary conditions imposed along one of the spatial directions. This abstract cylinder contains two edges and already allows to demonstrate the existence of helical edge states induced by the SOI. The corresponding energy spectrum (c.f.~Fig.~\ref{fig:spectrum:cylinder}a) depicts several edge-state channels: for each energy value within the bulk energy-gap, there exists a single time-reversed (or Kramers) pair of eigenstates localized on each edge of the lattice. The conservation of TRS prevents the mixing of this couple of states by small external perturbations and scattering from disorder~\cite{Kane2005,Kane2005bis}.

The helical edge states characterizing the QSH phase are topologically protected against external perturbations. Their property can be quantified by looking at the $\text{Z}_2$-index $\nu$~\cite{Kane2005}. This topological invariant characterizes the eigenstates defined in the bulk | it is defined on a 2-torus, in direct analogy with the Chern numbers introduced in the quantum Hall effect. Following Ref.~\cite{Franz:2010}, we have calculated this Z$_2$-index $\nu$ using the inversion symmetry of the lattice~\cite{Fu2007}. We obtain that SOI opens a spectral gap characterized by the index $\nu=1$, therefore classifying the Lieb lattice as a quantum spin-Hall insulator.

In the absence of spin-mixing perturbations, the Z$_2$ index is related to the spin Chern number $n_\sigma$~\cite{Sheng2006,Fukui2007}, through the simple relation  $\nu=n_\sigma\text{mod}\,2$, where $n_\sigma=(N_{\uparrow}-N_{\downarrow})/2$ and $N_{\uparrow, \downarrow}$ represent the Chern numbers associated to the individual spins. Using the numerical method of Ref.~\cite{fukui:2005}, we have obtained $n_\sigma=1$ in agreement with the above result.

It is interesting to extend the analysis above by considering the more realistic open geometry, i.e. a finite piece of Lieb lattice, thus characterized by a unique edge. We have solved the TB problem for a Lieb lattice of $39 \times 39$ sites with realistic straight edges. This yields a discrete energy spectrum extending in the range $E \in [-2.8 t , 2.8 t]$. In the vicinity of the CNP, within a range corresponding to $\Delta_\text{gap}=4\  t_\text{SO}$, the eigenvalues correspond to eigenstates that are localized at the edge of the system. This result is illustrated in Fig.~\ref{fig:spectrum:cylinder}b, where the amplitude $\vert \psi_{\uparrow} (x,y) \vert ^2$ is drawn for a particular edge-state at $E=0.5 t$. Note that this coincides with $\vert \psi_{\downarrow} (x,y) \vert ^2$. Using this geometry, one can verify the helical property of these edge-states by computing their associated current, which for spins $\sigma$ can be expressed as
%
\begin{align}
j_x^{\sigma}(m,n)&= -\ii  \ell_0 \biggl ( \psi^*_{\sigma} (m+1, n) \, \mathcal{U}_x \, \psi_{\sigma} (m,n) \\
& + \sqrt{2} \psi^*_{\sigma} (m+1, n+1) \, \mathcal{D} \psi_{\sigma} (m,n)   \,    - \text{h.c}  \biggr) , \notag \\
j_y^{\sigma}(m,n)&= -\ii  \ell_0 \biggl ( \psi^*_{\sigma} (m, n+1) \, \mathcal{U}_y \, \psi_{\sigma} (m,n)  \\
&+\sqrt{2} \psi^*_{\sigma} (m+1, n+1) \, \mathcal{D} \, \psi_{\sigma} (m,n)     - \text{h.c} \biggr ). \notag
\end{align}
%
%
Here $\psi_{\sigma} (m,n)= ( \psi_{\sigma} (m,n, \text{H}), \psi_{\sigma} (m,n, \text{A}) , \psi_{\sigma} (m,n, \text{B}) )$ and
%
%
\begin{subequations}
\begin{align}
\mathcal{U}_x &=\begin{pmatrix}
0 &0 &0 \\
t \mathbb{I}_2 &0 &- \ii t_\text{SO} \sigma \\
0 &0 &0
\end{pmatrix}, \\
 \mathcal{U}_y & =\begin{pmatrix}
0 &0 & t \mathbb{I}_2 \\
0 &0 & - \ii t_\text{SO} \sigma \\
0 &0 &0
\end{pmatrix},\\
\mathcal{D}& =\begin{pmatrix}
0 &0 & 0 \\
0 &0 & \ii t_\text{SO} \sigma \\
0 &0 &0
\end{pmatrix}.
\end{align}
\end{subequations}
%
%
We have verified that the currents $\boldsymbol{j}^{\uparrow}(m,n)$ and $\boldsymbol{j}^{\downarrow}(m,n)$ associated to the edge-states yield two vector fields circulating along the edge in opposite directions.

%
%
%
%

%
%
\section{Low-energy fermions: the quasirelativistic regime\label{LWA}}

In this Section, we focus on the the properties of the Lieb lattice for non-interacting fermions at low energy, i.e. close to the CNP. We perform a long wavelength approximation to the Schr\"odinger equation underlying the TB Hamiltonian
which consists in expressing the spatial part of the wave function as the product of a fast-varying part times a slow-varying part. Within this approximation, the wave function can be written as
%
%
\begin{equation}\label{eq:wf:lwa}
\Psi_\alpha(\bm{R}_\alpha) \propto \ee^{\ii \bm{k}\cdot\bm{R}_\alpha} F_\alpha(\bm{R}_\alpha)\,,
\end{equation}
%
%
where $\alpha\in\{\text{A,B,H}\}$ and $\bm{R}_\alpha$ is the lattice site coordinate. We substitute this wave function into the Schr\"odinger equation and expand the slow-varying part as
%
%
\begin{equation*}\label{lwa}
F_\alpha(\bm{R}_{\alpha'}\pm\bm{d}_j)\simeq F_\alpha(\bm{R}_{\alpha'}) \pm \bm{d}_j \cdot \bm{\nabla}_{\bm{r}} \left.F_\alpha(\bm{r})\right|_{\bm{r}=\bm{R}_{\alpha'}} + \mathcal{O}(|\bm{d}|^2)\,.
\end{equation*}
%
%
Collecting all terms, we are left with
%
%
\begin{equation}\label{lwaHam}
\tilde{\mathcal{H}}= v_\text{F} \bm{\Sigma}\cdot \bm{p} \,,
\end{equation}
%
%
where $v_\text{F}=2\ell_0t$ is the Fermi velocity, and $\bm{p}=(p_x,p_y,0)=-i\hbar(\partial_x,\partial_y,0)$. Here the pseudo-spin matrices are defined as
%
%
\begin{equation}\label{eq:gm}
\Sigma_x=
\begin{pmatrix}
0 & 1 & 0 \\
1 & 0 & 0 \\
0 & 0 & 0
\end{pmatrix},\
\Sigma_y=  \begin{pmatrix}
0 & 0 & 0 \\
0 & 0 & 1 \\
0 & 1 & 0
\end{pmatrix},\
\Sigma_z=\begin{pmatrix}
0 & 0 & -\ii \\
0 & 0 & 0 \\
\ii & 0 & 0
\end{pmatrix}\,.
\end{equation}
%
%
These matrices fulfill the algebra of the angular momentum $[\Sigma_i,\Sigma_j]=\ii \epsilon_{ijk}\Sigma_k$
and form a 3-dimensional representation of SU(2). However, contrary to the Pauli matrices, they do not form a Clifford algebra, \emph{i.e.}, \ $\{\Sigma_i,\Sigma_j\}\neq 2 \delta_{i,j} \mathbb{I}_3$. Therefore, while Eq.~\eqref{lwaHam} describes electrons with a linear energy spectrum, it does not represent a Dirac Hamiltonian.
By introducing a rotation operator around the $z$ axis defined by $\mathcal{D}_z(\phi) = \exp \left( -\ii \Sigma_z \phi \right)$, a generic state $\ket{\alpha}$ is transformed into itself by $\mathcal{D}_z(2\pi)\ket{\alpha}\to\ket{\alpha}$, implying that the pseudo-spin $\bm{\Sigma}$ describes an integer spin $S=1$.

\subsection{Spin-orbit interaction}

Within the long wavelength approximation we can also express the intrinsic SOI introduced in the previous Section. This term reads
%
%
\begin{equation}\label{hsoi:lwa}
\tilde{\mathcal{H}}_\text{SO} = \Delta_\text{SO} \Sigma_z \otimes \sigma_z ,
\end{equation}
%
%
where $\Delta_\text{SO}$ is the effective spin-orbit coupling strength and $\sigma_z$ is a Pauli matrix.
The energy spectrum  can be computed in this regime and reads
%
%
\begin{subequations}\label{soi:lwa}
\begin{align}
\tilde{\varepsilon}_0^\text{(SO)} &= 0 \\
\tilde{\varepsilon}_\pm^\text{(SO)} &= \pm \sqrt{v_\text{F}^2 |\bm{k}|^2 + \Delta_\text{SO}^2}\,.
\end{align}
\end{subequations}
%
%
This is two-fold degenerate, with degeneracy corresponding to spin-up and spin-down. \comment{The eigenstates for the case of spin-up reads
%
%
\begin{subequations}\label{eigenvectors:soi}
\begin{align}
\bm{v}_0 & = \begin{pmatrix}
	-t \cos(\bm{d}_\text{A}\cdot\bm{k}) \\ 2 \ii \Delta_\text{SO}  \sin (\bm{d}_\text{A}\cdot\bm{k}) \sin
   (\bm{d}_\text{B}\cdot\bm{k})\\ t \cos (\bm{d}_\text{A}\cdot\bm{k})
   \end{pmatrix} , \\
\bm{v}_\pm & =  \\
& \hspace{-0.6cm}\pm\!\!\nonumber\begin{pmatrix}
\varepsilon_\pm  \cos (\bm{d}_\text{A}\!\cdot\!\bm{k})-4 \ii \Delta_\text{SO}  \sin (\bm{d}_\text{A}\!\cdot\!\bm{k}) \sin
   (\bm{d}_\text{B}\cdot\bm{k}) \cos (\bm{d}_\text{B}\cdot\bm{k})\\
   t (\cos (\bm{d}_\text{A}\!\cdot\!\bm{k})+\cos (\bm{d}_\text{B}\!\cdot\!\bm{k})) \\
   \varepsilon_\pm  \cos (\bm{d}_\text{A}\!\cdot\!\bm{k})+4 \ii \Delta_\text{SO}  \sin
   (\bm{d}_\text{A}\!\cdot\!\bm{k}) \cos (\bm{d}_\text{A}\!\cdot\!\bm{k}) \sin (\bm{d}_\text{B}\!\cdot\!\bm{k})
\end{pmatrix}\!.
\end{align}
\end{subequations}
%
%
}

\subsection{Landau levels in a uniform magnetic field}

The long wavelength approximation also allows to compute the Landau levels that arise in the presence of the uniform magnetic field.
The system Hamiltonian reads
%
%
\begin{equation}\label{eq:ham:landau}
  \tilde{\mathcal{H}}_B = v_\text{F}\,\bm{\Sigma}\cdot\left(\bm{p}-\frac{e}{c}\bm{A}\right)\,,
\end{equation}
%
%
where  $\mathbf{A}$ is the
vector potential associated with the magnetic field
$\bm{B}=(\bm{\nabla}\times\bm{A})$
perpendicular to the lattice plane.

We solve the Schr\"odinger equation in the Landau gauge with $\mathbf{A}=(-By,0,0)$.
Further, we make the Ansatz $\Psi=\psi(y)\exp(\ii kx)$. Introducing a $k$--dependent shift
in the $y$--coordinate, $\sqrt{B}\yy=By+k$ we are left to solve a system
of coupled linear differential equations
%
%
\begin{subequations}
\begin{align}
   \yy\psi_\text{H}(\yy)&=\tilde\eps\psi_\text{A}(\yy)\\
   \yy\psi_\text{A}(\yy)-\ii\psi_\text{B}'(\yy)&=\tilde\eps\psi_\text{H}(\yy)\\
   -\ii\psi_\text{H}'(\yy)&=\tilde\eps\psi_\text{B}(\yy)
\end{align}
\end{subequations}
%
%
for the three components of $\psi(\yy)$. Here $\eps=E/(\hbar v_\text{F}\sqrt{B})$ is the
rescaled eigenenergy.
The Landau levels at non-zero energy are given by
%
%
\begin{equation}\label{eq:LL_finiteE}
   \psi_{\pm,\mathfrak{n}}(\yy) =
   \begin{pmatrix}
     (\sqrt{\mathfrak{n}}\phi_{\mathfrak{n}-1}+\sqrt{\mathfrak{n}+1}\phi_{\mathfrak{n}+1})/\sqrt{2} \\
     \pm\sqrt{2\mathfrak{n}+1}\phi_\mathfrak{n} \\
     -i(\sqrt{\mathfrak{n}}\phi_{\mathfrak{n}-1}-\sqrt{\mathfrak{n}+1}\phi_{\mathfrak{n}+1})/\sqrt{2}
   \end{pmatrix}
\end{equation}
%
%
with corresponding eigenvalues $\tilde\eps_{\pm,\mathfrak{n}}=\pm\sqrt{2\mathfrak{n}+1}$ and $\mathfrak{n}$ integer.
Here the $\phi_\mathfrak{n}$ for $\mathfrak{n}\ge0$ are the eigenfunctions of the one-dimensional harmonic oscillator,
%
%
\begin{equation}
    \phi_{\mathfrak{n}}(\yy) = \frac{1}{\sqrt{2^\mathfrak{n}\pi^{1/2}\mathfrak{n}!}} h_{\mathfrak{n}}(\yy)e^{-\yy^2/2},
\end{equation}
%
%
where $h_{\mathfrak{n}}$ denotes the Hermite polynomial of order $\mathfrak{n}$, while we define $\phi_{-1}\equiv 0$.
In addition to the eigenfunctions (\ref{eq:LL_finiteE}) there are two different types of solutions at energy $\tilde\eps_{0,\mathfrak{n}}=0$.
The first of these is related to the flat band at zero magnetic field (\ref{spec:zero}) and reads
%
%
\begin{equation}\label{eq:LL_zeroE}
   \psi_{0,\mathfrak{n}}(\yy) = \frac{1}{\sqrt{2}}
   \begin{pmatrix}
     \sqrt{\mathfrak{n}+1}\phi_{\mathfrak{n}-1}-\sqrt{\mathfrak{n}}\phi_{\mathfrak{n}+1} \\
     0 \\
     -\ii(\sqrt{\mathfrak{n}+1}\phi_{\mathfrak{n}-1}+\sqrt{\mathfrak{n}}\phi_{\mathfrak{n}+1})
   \end{pmatrix},
\end{equation}
%
%
where $\mathfrak{n}>0$. The second type of zero-energy Landau level
is given by
%
%
\begin{equation}\label{eq:LL_zeroE_B}
   \psi_{0}^{(0)}(\yy)=\frac{1}{\sqrt{2}}
   \begin{pmatrix}
     \phi_{0} \\
     0 \\
     \ii\phi_{0}
   \end{pmatrix}
\end{equation}
%
%
Note that $\psi_0^{(0)}(\yy)$ is not a generalization of (\ref{eq:LL_zeroE}) to the case $\mathfrak{n}=0$.

\subsection{Spectrum with magnetic field and spin-orbit interaction}

Now we turn to the effects of finite SOI on the Landau levels obtained above. The energies $\tilde\eps_{\alpha,\sigma,\mathfrak{n}}$ (with $\alpha=\{0,\pm\}$ and $\sigma=\{\uparrow,\downarrow\}$) of the Landau levels are the three solutions of
%
%
\begin{eqnarray}
    \frac{\sqrt{\mathfrak{n}}}{\tilde\eps_{\alpha,\sigma,\mathfrak{n}}-\sigma\Delta_\SO}+\frac{\sqrt{\mathfrak{n}+1}}{\tilde\eps_{\alpha,\sigma,\mathfrak{n}}+\sigma\Delta_\SO}&=&\tilde\eps_{\alpha,\sigma,\mathfrak{n}}\,.
\end{eqnarray}
%
%
The corresponding wave functions read
%
%
\begin{eqnarray}
\label{eq:LL_finiteSOI}
   \psi_{\alpha,\sigma,\mathfrak{n}}&=&
   \frac{\sqrt{\mathfrak{n}}}{\tilde\eps_{\alpha,\sigma,\mathfrak{n}}-\sigma\Delta_\SO}\phi_{\mathfrak{n}-1}
   \begin{pmatrix}
     1\\
     0\\
     -\ii
   \end{pmatrix}
   +\phi_\mathfrak{n}
   \begin{pmatrix}
     0\\
     \sqrt{2}\\
     0
   \end{pmatrix}
\nonumber\\&&
   +\frac{\sqrt{\mathfrak{n}+1}}{\tilde\eps_{\alpha,\sigma,\mathfrak{n}}+\sigma\Delta_\SO}\phi_{\mathfrak{n}+1}
   \begin{pmatrix}
     1\\
     0\\
     \ii
   \end{pmatrix}.
\end{eqnarray}
%
%

In the case of weak SOI, i.e. $\Delta_\SO\ll1$, the Landau levels are given by
%
%
\begin{eqnarray}
    \tilde\eps_{\pm,\sigma,\mathfrak{n}}&=&\pm\sqrt{2\mathfrak{n}+1}-\frac{\sigma\Delta_\SO}{4\mathfrak{n}+2}+{\cal O}(\Delta_\SO^2) \\
    \eps_{0,\sigma,\mathfrak{n}}&=&\frac{\sigma\Delta_\SO}{2\mathfrak{n}+1}+{\cal O}(\Delta_\SO^2)
\end{eqnarray}
%
%
Consequently, the main effect of finite SOI is to lift the spin degeneracy, with a
level separation that decreases with growing Landau level index $\mathfrak{n}$. Moreover, the former highly degenerate
zero energy levels, c.f. Eqs.~(\ref{eq:LL_zeroE}) and (\ref{eq:LL_zeroE_B}), are now split into a family of flat bands at energies $\eps_{0,\sigma,\mathfrak{n}}$.\comment{\bcom{Missing plots of the Landau levels + Landau levels modified by SOI}}

\section{Optical lattice realization}

\begin{figure}[!t]
\centering
\includegraphics[width=\columnwidth]{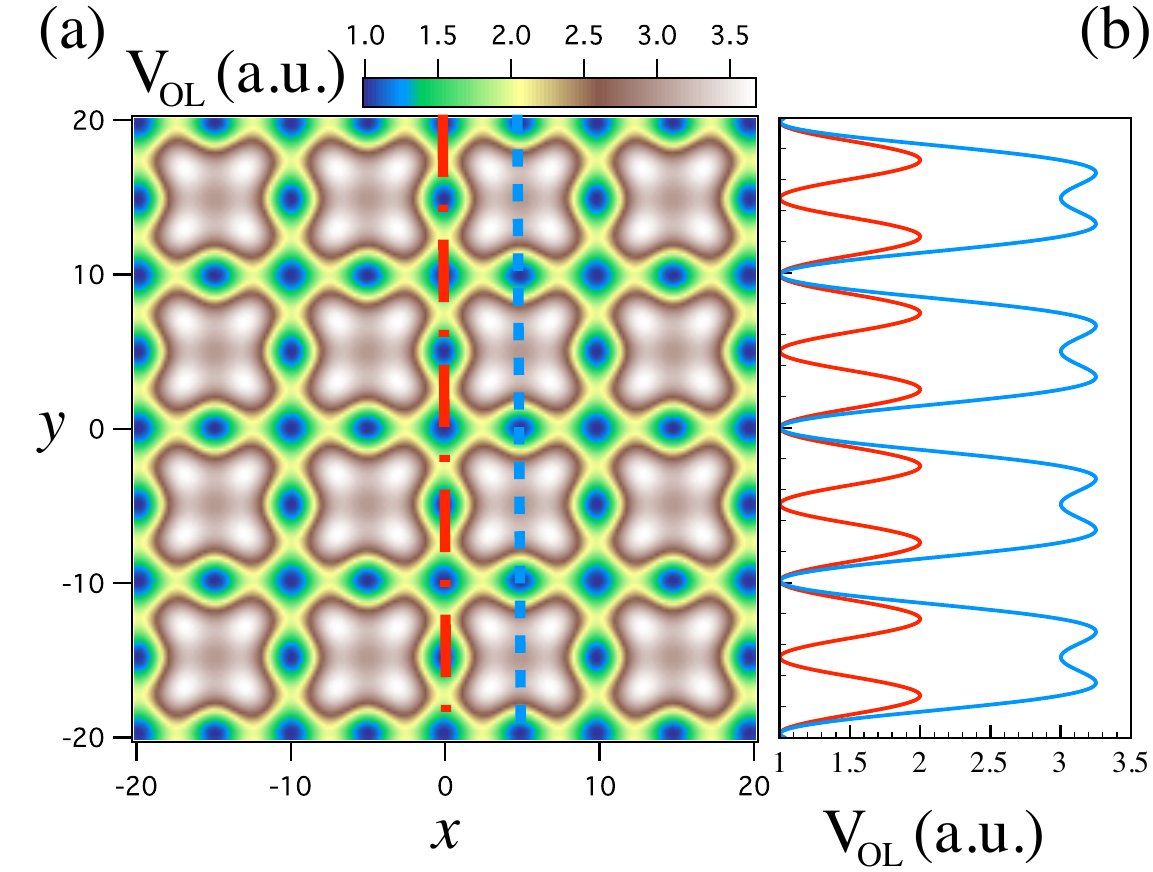}
\caption{\label{fig:optical:lattice} (Color online) (a) Distribution of the
laser field intensity for generating a Lieb lattice.  (b) Two cuts through (a) corresponding to the B and H sites (red line) and A sites (blue line)}
\end{figure}

Experimentally, the Lieb lattice can be realized as an optical lattice created by  six counter propagating
pairs of laser beams. Four pairs are aligned along the $x$ and $y$ directions, two with wavelength $\lambda =\ell_0$ and two with wavelength $\lambda =\ell_0/2$. Finally, it requires two other laser pairs with a direction of $\pm45^\circ$ with respect to the $x$-axis. A detailed procedure leading to this choice of laser configuration has been discussed in Refs.~\cite{schen:2010,apaja:2010}. The potential profile is given by the field
%
%
\begin{align}\label{v:ol}
V_\text{OL}(x,y) = &  V_0 \left[ \sin^2(kx)+\sin^2(ky)\right] \\
& + V_1 \left[ \sin^2(2kx)+\sin^2(2ky)\right]\nonumber \\
& + V_2 \left[ \cos^2\left(k \frac{x+y}{2} \right)+\cos^2\left(k \frac{x-y}{2} \right)\right] , \nonumber
\end{align}
%
%
with $V_0=V_1=2V_2$ and $k=\pi/\lambda$, which is depicted in Fig.~\ref{fig:optical:lattice}.
It is apparent that the  hopping probability between A and B
sites is exponentially small compared to the hopping probability
between neighboring H and A/B pairs.

%
%
%
%

\subsection{Simulation of the U(1) synthetic gauge field and quantum Hall phases}

Recently, synthetic U(1) magnetic fields for cold neutral atoms have been proposed~\cite{Jaksch2003,gerbier:2010,Dalibard2010} and experimentally realized~\cite{lin:2009}. In such setups, atoms reproduce the dynamics of charged particles subjected to a uniform magnetic field and can effectively show quantum Hall phases. Several methods can be used in order to simulate the
Hofstadter model~\cite{Hofstadter1976} with these systems. These methods are generally based on the fact that the Peierls phases can be engineered by external optical~\cite{Jaksch2003,gerbier:2010} or magnetic~\cite{goldman:2010} fields. These electromagnetic fields can indeed induce hopping between neighboring lattice sites, when the latter host atoms in different internal states, say $\vert g \rangle$ and $\vert e \rangle$. More precisely, the external fields trigger (Raman) couplings between these internal states, resulting in a NN-hopping amplitude
\begin{equation}
t_{g,e} e^{i \theta (\boldsymbol{x}_g)} \propto \int  w^* (\boldsymbol{x}-\boldsymbol{x}_e) \Omega_{g , e}  w (\boldsymbol{x}-\boldsymbol{x}_g) d^3 x,
\label{raman}
\end{equation}
where the Rabi frequency $\Omega_{g , e} $ typically includes space-dependent phase factors and where we suppose the states $\vert g \rangle$ and $\vert e \rangle$ to be trapped in neighboring sites ~\cite{Jaksch2003,gerbier:2010}. Here, the Peierls phase $\exp (i \theta (\boldsymbol{x}_g))$ is directly related to the coupling laser's wave vector.

The gauge field \eqref{gaugeu1}, which leads to the Peierls phase \eqref{peierls} can be readily engineered in an optical lattice experiment by exploiting these methods~\cite{Jaksch2003,gerbier:2010}. From Fig.~\ref{fig:lattice}, it is clear that the $x$-dependent assisted-hopping involves nearest-neighbors and occurs along the $y$ direction only, i.e. between B and H sites. In this sense, the phases $\theta(m)=\pi \Phi m$ can be realized on the optical Lieb lattice by extending the methods envisaged for the standard square lattice (cf. Ref.~\cite{Jaksch2003,gerbier:2010,Dalibard2010}): In the Lieb lattice case, one should trap two internal states alternatively along the $y$-direction: the B-sites (resp. H-sites) should host an atom in the internal state $\vert g \rangle$ (resp. $\vert e \rangle$). Coupling these states with external fields should then induce hoppings of the form \eqref{raman}, resulting in the space-dependent Peierls phase \eqref{peierls}. Note that for generating a magnetic flux $\Phi$ per plaquette, a double phase $\theta_{\square}=2 \theta=2 \pi \Phi m$ is required for the square lattice compared to the Lieb lattice.

In the previous sections, we have discussed the existence of quantum Hall phases in a fermionic Lieb lattice subjected to a uniform magnetic field. In order to produce these phases in a cold atom experiment, one should engineer a U(1) gauge field for \emph{fermionic} atoms. As already discussed in Ref.~\cite{goldman:2010}, most of the schemes generating gauge fields for bosons use Raman transitions to couple the internal states, and would lead to high spontaneous emission rates for fermionic atoms. Therefore realizing (integer) quantum Hall states would require alternative methods~\cite{Gerbier2009,mazza1,mazza2}.  Such a proposal was introduced in Ref.~\cite{goldman:2010} and uses radio-frequency magnetic fields produced by a set of current-carrying wires. The latter are periodically spaced on an atom chip and drive transitions between several internal states of $^{6}$Li  fermionic atoms. These effective ``Raman transitions" lead to assisted hopping \eqref{raman} and can be tuned in order to produce the desired Peierls phases. In order to engineer the U(1) gauge field  \eqref{gaugeu1}, one can simplify the method initially proposed in Ref.~\cite{goldman:2010} (which leads to the creation of SU(2) gauge fields) and consider transitions between two internal states of $^{6}$Li, e.g. $\ket{g_1} = \ket{F\!=\!1/2,m_F\!=\!1/2}$ and $\ket{e_1} = \ket{3/2,1/2}$. We stress that this practical scheme can  be directly generalized to the Lieb lattice in order to generate the phases $\theta(m)=\pi \Phi m$ accompanying the hopping between neighboring H and B sites.

%
%
\begin{figure}[tbp]
\begin{center}
\includegraphics[width=0.7\columnwidth]{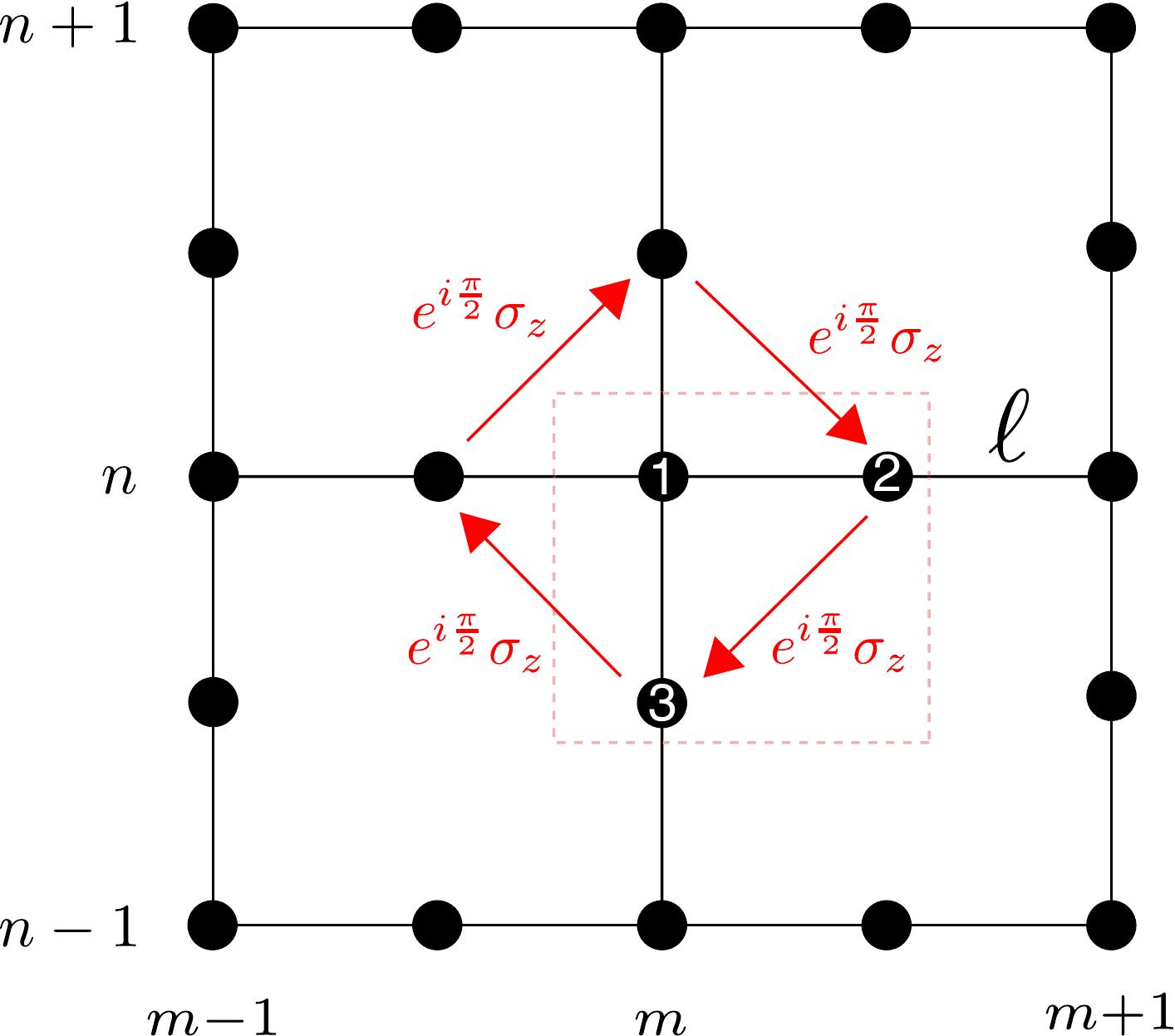}
\end{center}
\caption{\label{fig:Peierls:phases} The Lieb lattice with the intrinsic spin-orbit coupling. This coupling is equivalent to a non-Abelian gauge field $\bm{A}$ leading to spin-dependent Peierls phases, as indicated by red arrows.}
\end{figure}
%
%

\subsection{Simulation of the SU(2) gauge field with neutral atoms}

%
%
\begin{figure}[tbp]
\begin{center}
\includegraphics[width=1.2\columnwidth]{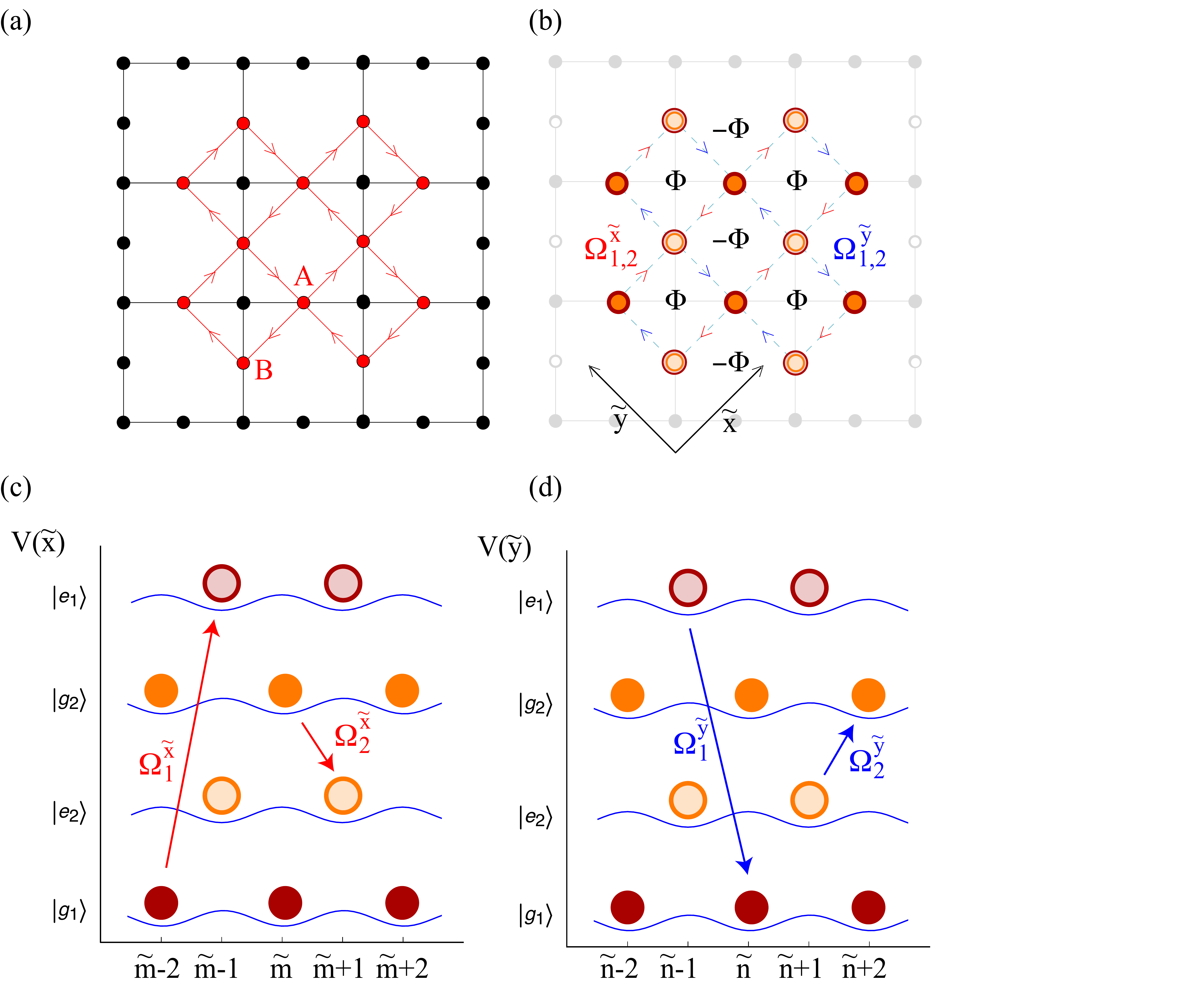}
\end{center}
\caption{\label{fig:SU2} (a) The Lieb lattice with the intrinsic spin-orbit coupling. The Peierls phases (red arrows) are defined on the square sublattice formed by the  A and B sites. (b) Laser assisted-hoppings (arrows), associated couplings and synthetic fluxes felt by the up-spin component. Four states  $e_{1,2}$ and $g_{1,2}$ are trapped in state-dependent potentials and are represented by filled or empty orange or red disks (cf. figures below). (c)-(d) State-dependent potentials and couplings along the $\tilde{x}$ and $\tilde{y}$ directions.}
\end{figure}
%
%

Here, we describe a practical scheme to simulate the intrinsic SOI term in a fermionic Lieb lattice. This coupling is equivalent to a SU(2) gauge field and could therefore be engineered in a multi-component atomic system through state-dependent Peierls phases (cf. Fig.~\ref{fig:Peierls:phases}). Our proposal is based on the observation that the spin-dependent NNN-hoppings are equivalent to \emph{NN}-hoppings defined on the square sublattice formed by the  A and B sites only~\cite{Liu:2010} (cf. Fig.~\ref{fig:SU2} (a)). Consequently, generating the SOI reduces to the simple problem of engineering the Peierls phases $\ee^{\ii \pi /2} \sigma_z$ on a \emph{rotated} square lattice, which we now label using the notations $\tilde{m}$ and $\tilde{n}$ (cf. Fig.~\ref{fig:SU2} (b)). Obviously, the subtlety relies in the orientation of these phases: the phases are positive for a particle hopping respectively clockwise and anti-clockwise in neighboring plaquettes (cf. Fig.~\ref{fig:SU2} (a)). Note that this important fact naturally leads to a staggered magnetic field, with fluxes $\pm \Phi$, for each spin component (cf. Fig.~\ref{fig:SU2}b). We note that, in order to reproduce such a staggered field, one can simply exploit the fact that the hopping induced by Raman transitions between the internal states $g$ and $e$ is such that $(t e^{i \theta (\boldsymbol{x}_g)})_{e,g}=(t e^{i \theta (\boldsymbol{x}_g)})_{g,e}^*$ \cite{Gerbier2009}. \\

Let us now describe a feasible and concrete scheme to synthesize the SOI term \eqref{eq:SO} in a cold-atom experiment. Our proposal requires four states $e_{1,2}$ and $g_{1,2}$ and external fields producing Raman transitions in both the $\tilde{x}$ and $\tilde{y}$ directions (cf. Figs.~\ref{fig:SU2} (a)-(d)). Such states can be chosen as being four internal states \mbox{$\ket{F,m_F}$} of $^{6}$Li, e.g. $\ket{g_1} =$\mbox{$\ket{1/2,1/2}$}, $\ket{g_2} =$\mbox{$\ket{3/2,-1/2}$}, $\ket{e_1} =$\mbox{$\ket{3/2,1/2}$}, and $\ket{e_2} =$\mbox{$\ket{1/2,-1/2}$}~\cite{goldman:2010}. First, one needs to trap these states in state-dependent lattices \cite{Osterloh2005,Gerbier2009,goldman:2010} along the $\tilde{x}$ and $\tilde{y}$ directions (cf.  Figs.~\ref{fig:SU2}c-d). Then external fields should drive Raman transitions between these states, with the corresponding Rabi frequencies
\begin{align}
&\Omega_{g_1 , e_1}^{\tilde{x}}=\Omega_1^{\tilde{x}}, \qquad \Omega_{g_2 , e_2}^{\tilde{x}}=\Omega_2^{\tilde{x}} , \notag \\
&\Omega_{e_1 , g_1}^{\tilde{y}}=\Omega_1^{\tilde{y}}, \qquad \Omega_{e_2 , g_2}^{\tilde{y}}=\Omega_2^{\tilde{y}} ,
\label{omega}
\end{align}
as indicated by arrows in Figs.~\ref{fig:SU2}c-d. At this point, we emphasize that the Rabi frequencies are controlled by the coupling lasers and that they are chosen to be different for transitions driven along the $\tilde{x}$ and $\tilde{y}$ directions. Note that these Rabi frequencies typically contain phase factors depending on the coupling lasers wave vectors \cite{Jaksch2003}. Moreover the Rabi frequencies associated to the opposite transitions are simply given by $\Omega_{e_j , g_j}^{\tilde{\mu}}=(\Omega_{g_j , e_j}^{\tilde{\mu}})^*$, where $\tilde{\mu}=\tilde{x},\tilde{y}$. Therefore, the hopping amplitudes along a given direction are accompanied by Peierls phases with alternating signs \cite{Gerbier2009}. This leads to a staggered magnetic field for each spin component, with fluxes $\pm \Phi$, as illustrated in Fig.~\ref{fig:SU2}b.\\

Now, the SOI term \eqref{eq:SO} requires a specific configuration of these Rabi frequencies. The associated SU(2) gauge field is proportional to $\sigma_z$, which is simply achieved by imposing the constraint $\Omega_{1}^{\tilde{\mu}}=(\Omega_{2}^{\tilde{\mu}})^*$ (i.e. the coupling lasers should be characterized by opposite wave vectors). Besides, the desired gauge field is associated to constant Peierls phases (i.e. $\exp( \pm i \pi /2)$), which further requires that the Rabi frequencies do not depend on the variables $\tilde{x},\tilde{y}$ and also obey the relation $\Omega_{1,2}^{\tilde{x}}=\Omega_{1,2}^{\tilde{y}}$ (or equally $\Omega_{g_j , e_j}^{\tilde{x}}=(\Omega_{g_j , e_j}^{\tilde{y}})^*$, using the definitions \eqref{omega}). \\

We stress that this concrete scheme leads to the SOI studied in the previous sections and that it should open a QSH gap in an atomic setup. In order to observe the QSH phases induced by such a synthetic SOI in a cold atom experiment, these non-trivial Peierls phases need to be engineered in a fermionic lattice. Again, one can consider the atom-chip proposal of Ref.~\cite{goldman:2010}: different sets of wires, aligned along $\tilde{x}$ and $\tilde{y}$, should trap the states $e_j$ and $g_j$ alternatively in both directions, as illustrated in Fig.~\ref{fig:SU2}c-d. Additional ``Raman wires" should then trigger RF transitions and couple the states, producing the induced-hopping and associated phases described above. Another possibility would be to apply the superlattice methods of  Refs.~\cite{Gerbier2009,mazza1,mazza2} to the Lieb lattice. Once the gauge field is synthesized, this setup needs to be superimposed with a state-independent Lieb lattice yielding the desired total Hamiltonian $\mathcal{H}=\mathcal{H}_{0}+\mathcal{H}_\text{SO}$. \\

Finally, we note that the scheme presented in this Section could be simplified in order to reproduce the Abelian Haldane model on the Lieb lattice~\cite{Haldane1988,Liu2010}. In this case, only two internal states $g$ and $e$ would be needed, instead of four. A realization of the Haldane model on the Lieb lattice would lead to integer quantum Hall states~\cite{Liu2010}.  

\subsection{Detection}

The TB Hamiltonian and its long wavelength
approximation are valid for a Lieb lattice
populated by single-component fermionic atoms, \emph{e.g.}
$^{40}$K or $^6$Li. In this case the atomic
collisions are negligible at low
temperature~\cite{Bloch2008}. From the experimental point
of view, time-of-flight imaging via light
absorption~\cite{koehl:2005} can be used in order to detect the
presence of massless fermions. The harmonic trap potential
$V(\mathbf r)=m\omega^2\mathbf{r}^2/2$ confining the fermionic
cold atom gas is ramped down slowly enough for the atoms to stay
adiabatically in the lowest band while their quasi-momentum is
approximatively conserved. Under these conditions, free fermions
expand with
ballistic motion and, from the measured absorption images, it is
possible~\cite{ho:2000,Umucalilar:2008} to reconstruct the initial reciprocal-space density profile
of the trapped gas. Then, the local
density approximation is typically well satisfied and the
local chemical potential can be assumed to vary with the radial
coordinate as $\mu(\mathbf{r}) = \mu_0 -V(\mathbf{r})$, where $\mu_0$ is the
chemical potential at the center of the trap.
For a system of cold atoms at temperature $T$, the atomic density in the bulk is uniquely
determined by the chemical potential
%
%
\begin{equation}\label{eq:twoelve}
\rho(\mu) = \frac{1}{\mathcal{S}_0} \sum_{\alpha} \int f(\mathbf{k},\alpha, \mu) \, d \mathbf{k} \,.
\end{equation}
%
%
Here $\mathcal{S}_0$ is the area of the
first Brillouin zone of the Lieb lattice, and
$f(\mathbf{k},\alpha, \mu)=[\exp[(E_\alpha(\mathbf{k})-\mu)/k_\text{B}T]+1]^{-1}$
is the Fermi distribution function, where $E_\alpha(\mathbf{k})$
is the energy spectrum of the Lieb lattice,
cf.\ Eq.~(\ref{eq:spectrum}).
%
%
\begin{figure}[!t]
\centering
\includegraphics[width=\columnwidth]{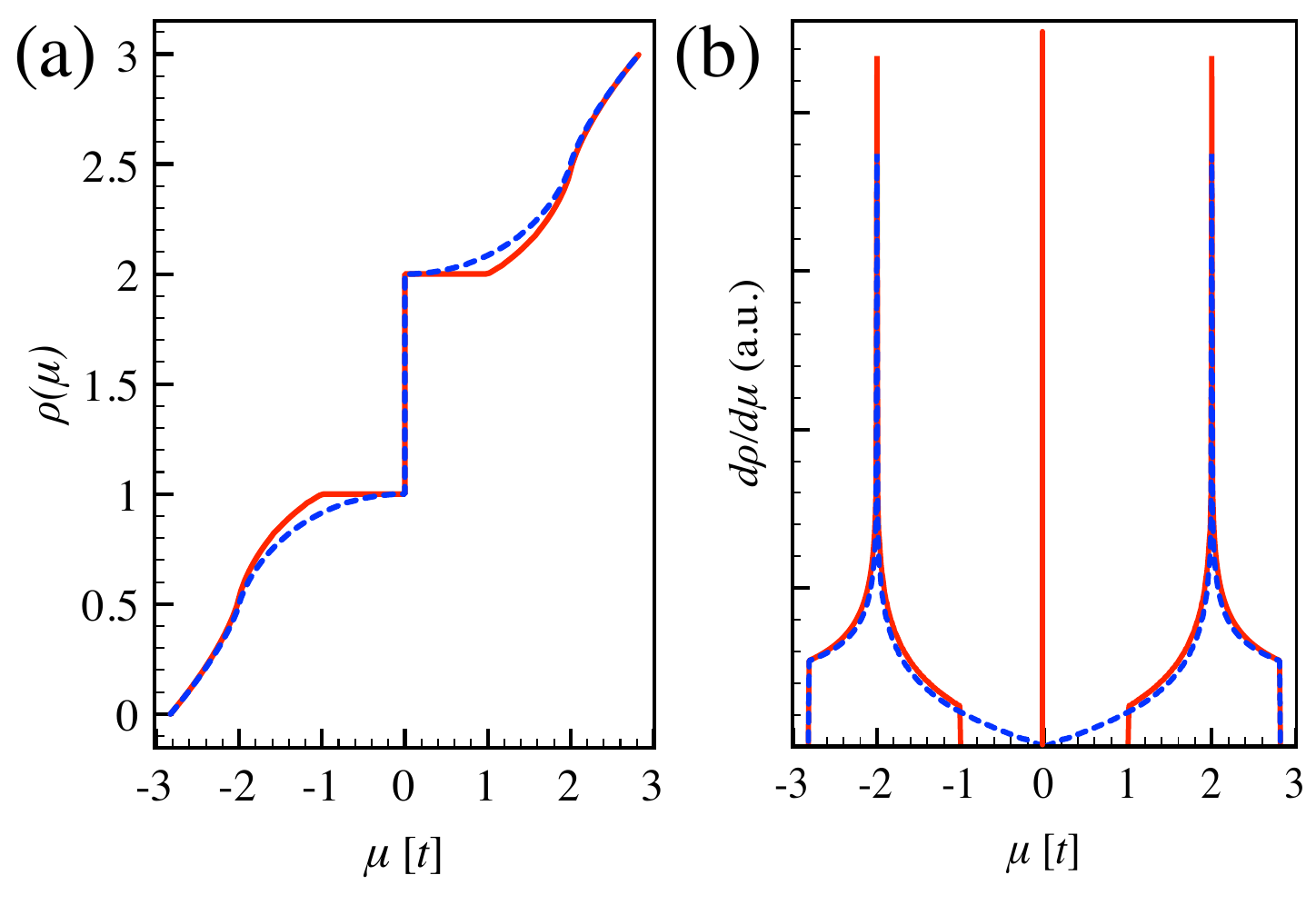}
\caption{\label{fig:density} (Color online) Panel (a) Atomic density as a function of the chemical potential. Panel (b) Density  of states as a function of the chemical potential.  The full line corresponds to $\Delta_{\text{SO}}=0.25t$ and the dashed line to $\Delta_{\text{SO}}=0$.}
\end{figure}
%
%
Figure~\ref{fig:density}a shows the atomic density $\rho$ for the bulk
as a function of the chemical potential $\mu$.
The contribution from the highly degenerate topological band (\ref{spec:zero}) manifests itself at $\mu=0$ as a sharp jump in the atomic density. This feature is specific to the Lieb lattice.
For small finite $\mu$ we note that $\rho$ evolves proportional to $\mu^2$, which
reflects the linear dispersion of massless fermions near the band center as well as particle-hole symmetry.
On the contrary, for values of $\mu$ close to the maximum or minimum of
the energy band, \emph{i.e.}\ far away from the band center (where the
long wavelength approximation can no longer be applied),
$\rho$ varies proportional to $\mu$. When we consider a finite value of the SOI, a finite gap appears in the energy spectrum. This coincides with the horizontal segments in Fig.~\ref{fig:density}a. Moreover, we can observe that the atomic density  in proximity of the SOI gap is not anymore a quadratic function of the chemical potential but is behaving linearly. This is a signature of the mass term introduced by the finite SOI. In Fig.~\ref{fig:density}b we report the density of states (DOS) for two values of $\Delta_{\text{SO}}$. Note the  gap opening for $\Delta_{\text{SO}}\ne 0$ and the existence of two robust van Hove singularities at $\mu= \pm 2$. \\

Let us stress the great similarity between the density and DOS obtained here for the Lieb lattice and those reported in Ref.~\cite{bercioux:2009} concerning the \ttt lattice. However, these two lattices differ in their topological order since the SOI opens a \emph{trivial} insulating gap in the \ttt lattice~\cite{bercioux:2010}. This fundamental difference can be emphasized by computing the density profile $\rho(\mu)$ for the realistic finite-size system: in this case, gapless edge states  contribute to the Lieb lattice atomic density and slightly tilt the SO plateaus. Note that this important effect, which is the direct signature of the topological phase, could only be observed for sufficiently small lattices (in which case the number of edge-states is not totally negligible compared to the number of bulk states). Another difficulty is that the edge-states could easily be destroyed by the harmonic potential: they should therefore be stabilized by sharp boundaries \cite{Stanescu2010} or by designed interfaces \cite{goldman:2010} . \\

Finally, the methods for detecting topological properties, such as quantized Hall conductivity \cite{Umucalilar:2008} and chiral edge-states \cite{Stanescu2010,Liu:2010,goldman:2010} have been discussed recently and could be easily generalized to our Lieb lattice setup.

\section{Summary}

We have investigated the fermionic properties of a face-centered square (Lieb) lattice. This peculiar system is characterized by the presence of a single Dirac cone at the center of the first Brillouin zone and a flat energy band at half filling. In particular we focused on the modification of this exquisite energy spectrum in the presence of an external magnetic field and a next-nearest-neighbor spin-orbit interaction. In the former case, we have shown the opening of multiple gaps due to the occurrence of Landau levels, which leads to the formation of  two Hofstadter butterflies separated by the robust flat band at zero energy. We have characterized the topological nature of these gaps by investigating  the IQHE. Inside the two Hofstadter butterflies, the Hall conductivity is quantized and each gap is characterized either by positive or negative values. Importantly, we find that the energy gaps separating the two butterflies have a trivial topological nature as they are characterized by zero Hall conductivity. This fact is a direct consequence of the flat band's trivial order. In the case of the spin-orbit interaction we have shown the opening of two symmetric gaps around the flat band. These are characterized by a non-trivial Z$_2$ topological phase leading to the quantum spin Hall effect. We have demonstrated the existence of helical edge states for an abstract cylindrical system, as well as for the more realistic finite lattice.

We have further investigated the Lieb lattice presenting a long wavelength approximation for the system Hamiltonian around the single Dirac cone. We have shown that the Hamiltonian can be expressed in a relativistic form -- similarly to the honeycomb lattice case -- which is characterized by a set of pseudo-spin matrices of size $3\times3$. These matrices fulfill the commutation relation of an angular momentum and describe a spin-1 particle.  Within the same approximation we have demonstrated that the spin-orbit interaction simply results in a mass term within the quasirelativistic Hamiltonian. Furthermore, we have inspected the properties of the Landau levels. In addition to the dispersion relation ruled by the square root of the Landau level index -- as in the case of the honeycomb lattice --  we have shown that  there are two competing Landau levels at zero energy, which are related to the lattice topology and to the symmetry class of the Hamiltonian operator.

This system and its associated properties could be engineered using fermionic cold atoms placed in an optical lattice resembling the Lieb lattice topology. We have proposed a method for implementing Abelian and non-Abelian synthetic gauge fields in order to simulate the presence of an external magnetic field and a next-nearest-neighbor spin-orbit interaction term. We have further shown that these synthetic fields trigger the opening of energy gaps while preserving the robust flat band at half-filling, properties which can be directly deduced from atomic density measurements. In particular, we emphasized that the Lieb lattice is very well suited to reproduce the intrinsic spin-orbit term introduced in Ref. \cite{Kane2005}, which in this case, can be simply decomposed into nearest-neighbour hoppings on a square sublattice.

\acknowledgments
NG thanks the F.R.S-F.N.R.S (Belgium) for financial support. DB is supported by the Excellence Initiative of the German
Federal and State Governments.


\begin{thebibliography}{99}






\bibitem{Klitzing1986} K. von Klitzing, Rev. Mod. Phys. \textbf{58}, 519 (1986).

\bibitem{Thouless1982} D. J. Thouless, M. Kohmoto, M. P. Nightingale, and M. den Nijs, Phys. Rev. Lett. \textbf{49}, 405 (1982).

\bibitem{Kohmoto1985} M. Kohmoto, Ann. Phys. \textbf{160}, 343 (1985).

\bibitem{Hatsugai1993} Y. Hatsugai, Phys. Rev. B \textbf{48}, 11851 (1993).

\bibitem{Qi2006} X.-L. Qi, Y.-S. Wu and S.-C. Zhang, Phys. Rev. B \textbf{74}, 045125 (2006).

\bibitem{Haldane1988} F. D. M. Haldane, Phys. Rev. Lett. \textbf{61},  2015 (1988).

\bibitem{Kane2005} C. L. Kane, and E. J. Mele, Phys. Rev. Lett.  \textbf{95}, 146802 (2005).

\bibitem{Kane2005bis} C. L. Kane, and E. J. Mele, Phys. Rev. Lett.  \textbf{95}, 226801 (2005).

\bibitem{Kane2006} C. L. Kane and E. J. Mele, Science \textbf{314}, 1692 (2006).

\bibitem{Bernevig2006} B. A. Bernevig and S.-C. Zhang, Phys. Rev. Lett. \textbf{96}, 106802 (2006).

\bibitem{Bernevig2006bis} B. A. Bernevig, T. L. Hughes and S.-C. Zhang, Science  \textbf{314}, 1757 (2006).

\bibitem{Qi2008} X.-L. Qi, T. L. Hughes, and S.-C. Zhang, Phys. Rev. B \textbf{78},  195424 (2008).

\bibitem{Avron1988} J. E. Avron, L. Sadun, J. Segert and B. Simon, Phys. Rev. Lett. \textbf{61},  1329 (1988).

\bibitem{Sheng2006} D. N. Sheng \emph{et al.}, Phys. Rev. Lett. \textbf{97}, 036808 (2006).

\bibitem{Fukui2007} T. Fukui and Y. Hatsugai, Phys. Rev. B \textbf{75}, 121403(R) (2007).

\bibitem{Guo2009} H.- M. Guo and M. Franz, Phys. Rev. B \textbf{80}, 113102 (2009).

\bibitem{Liu2010} G. Liu \emph{et al.}, Phys. Rev. A \textbf{82}, 053605 (2010).

\bibitem{Franz:2010} C. Weeks and M. Franz, Phys. Rev. B \textbf{82}, 085310 (2010).

\bibitem{Ruegg2010} A. Ruegg, J. Wen and G. A. Fiete, Phys. Rev. B \textbf{81}, 205115 (2010).

\bibitem{Stanescu2010} T. D. Stanescu, V. Galitski and S. Das Sarma, Phys. Rev. A \textbf{82}, 013608 (2010).

\bibitem{goldman:2010} N. Goldman, I. Satija,  P. Nikolic,  A. Bermudez, M. A. Martin-Delgado, M. Lewenstein and I. B. Spielman, Phys. Rev. Lett. \textbf{105},  255302 (2010).

\bibitem{bercioux:2010} D. Bercioux, N. Goldman, D.~F. Urban, arXiv:1007:2056, Phys. Rev. A (2011) \emph{in press}.

\bibitem{Sun2009} K. Sun, H. Yao, E. Fradkin and S. A. Kivelson, Phys. Rev. Lett. \textbf{103}, 046811 (2009).

\bibitem{Guo2009PRL} H.- M. Guo and M. Franz, Phys. Rev. Lett. \textbf{103}, 206805 (2009).

\bibitem{Fu2007} L. Fu, C. L. Kane and E.J. Mele, Phys. Rev. Lett. \textbf{98}, 106803 (2007).

\bibitem{Kargarian2010} M. Kargarian and G. A. Fiete, Phys. Rev. B \textbf{82}, 085106 (2010).

\bibitem{Tasaki2008} H. Tasaki, Eur. Phys. J. B  \textbf{64}, 365 (2008).

\bibitem{Lieb1989} E. H. Lieb, Phys. Rev. Lett. \textbf{62}, 1201 (1989).

\bibitem{Mielke1991} A. Mielke, J. Phys. A \textbf{24}, L74 (1991); \emph{ibid} \textbf{24},  3311  (1991);  \emph{ibid} \textbf{25},  4335  (1992).

\bibitem{Bergman2008} D. L. Bergman, C. Wu and L. Balents, Phys. Rev. B \textbf{78}, 125104 (2008).

\bibitem{Green2010} D. Green, L. Santos and C. Chamon, Phys. Rev. B \textbf{82}, 075104 (2010).

\bibitem{localization} J.~Vidal, R.~Mosseri, and B.~Dou\c cot, Phys. Rev. Lett. \textbf{81}, 5888 (1998); J.~Vidal, P.~Butaud, B.~Dou\c cot, and R.~Mosseri, Phys. Rev. B \textbf{64}, 155306 (2001); D. Bercioux, M. Governale, V. Cataudella and V.~M. Ramaglia, Phys. Rev. Lett. \textbf{93}, 056802 (2004); D. Bercioux, M. Governale, V. Cataudella and V.~M. Ramaglia, Phys. Rev. B \textbf{72}, 075305 (2005).

\bibitem{Dagotto} E. Dagotto, E. Fradkin and A. Moreo, Phys. Lett. B \textbf{172}, 383 (1986).

\bibitem{schen:2010} R. Shen, L.~B. Shao, B. Wang, and D.~Y. Xing, Phys. Rev. B \textbf{81}, 041410(R) (2010).



\bibitem{Lewenstein2007} M. Lewenstein, A. Sanpera, V. Ahufinger, B. Damski, A. Sen (De), and U. Sen, Adv. Phys. \textbf{56}, 243 (2007).

\bibitem{Bloch2008} I. Bloch, J. Dalibard, and W. Zwerger, Rev. Mod. Phys. \textbf{80}, 885 (2008).

\bibitem{apaja:2010} V. Apaja, M. Hyrk\"as, and M. Mannien, Phys. Rev. A  \textbf{82}, 041402 (2010).

\bibitem{grynberg:1993} G. Grynberg, B. Lounis, P. Verkerk, J.-Y. Courtois and C. Salomon, Phys. Rev. Lett.  \textbf{70}, 2249 (1993).

\bibitem{bercioux:2009} D. Bercioux, D.~F. Urban, H. Grabert  and W. H\"ausler, Phys. Rev. A \textbf{80}, 063603 (2009).

\bibitem{Liu:2010} X.-J. Liu,  C. Wu and J. Sinova, Phys. Rev. A \textbf{81}, 033622 (2010).

\bibitem{Lin2009} Y.-J. Lin \emph{et al.}, Phys. Rev. Lett. \textbf{102}, 130401 (2009).

\bibitem{Spielman2009} I. B. Spielman, Phys. Rev. A \textbf{79}, 063613 (2009).

\bibitem{Jaksch2003} D. Jaksch, and P. Zoller, New J. Phys. \textbf{5}, 56 (2003).

\bibitem{Ho2010} Tin-Lun Ho and Shizhong Zhang, arXiv:1007.0650v1.

\bibitem{Juze2010} G. Juzeliunas, J. Ruseckas and J. Dalibard, Phys. Rev. A \textbf{81}, 053403 (2010).

\bibitem{Spielman2010} I. Spielman, Conference at KITP, Santa Barbara (2010)

\bibitem{Wang2010b} Z. Wang, X.-L. Qi,  and S.-C. Zhang, New J. Phys. {\bf12} 065007 (2010).

\bibitem{Stanescu2008} T. D. Stanescu, B. Anderson and V. Galitski, Phys. Rev. A \textbf{78}, 023616 (2008).

\bibitem{Juze2005} G. Juzeliunas, P. Ohberg, J. Ruseckas and A. Klein, Phys. Rev. A \textbf{71}, 053614 (2005).

\bibitem{Dalibard2010} J. Dalibard \emph{et al.}, arXiv:1008:5378v1.

\bibitem{Lin2010}  Y.-J. Lin \emph{et al.}, arXiv:1008.4864v1.

\bibitem{Hofstadter1976} D. Hofstadter, Phys. Rev. B \textbf{14}, 2239 (1976).

\bibitem{Goldman2009} N. Goldman, A. Kubasiak, P. Gaspard, and M. Lewenstein, Phys. Rev. A. \textbf{79}, 023624 (2009).

\bibitem{Sorensen2004} A. S. S{\o}rensen,  E. Demler, and M. D. Lukin, Phys. Rev. Lett. \textbf{94}, 086803 (2004).

\bibitem{Merkl2008} M. Merkl, F. E. Zimmer, G. Juzeliunas and P. Ohberg, Europhys. Lett.  \textbf{83}, 54002 (2008).

\bibitem{Goldman2009bis} N. Goldman, A. Kubasiak, A. Bermudez, P. Gaspard, M.A. Martin Delgado, and M. Lewenstein, Phys. Rev. Lett. \textbf{103}, 035301 (2009).

\bibitem{Lim2008} Lih-King Lim, C. Morais Smith and  A. Hemmerich, Phys. Rev. Lett.  \textbf{100}, 130402 (2008).

\bibitem{Gunter2009} K. J. G\"unter, M. Cheneau, T. Yefsah, S. P. Rath and J. Dalibard, Phys. Rev. A \textbf{79}, 011604(R) (2009).

\bibitem{Osterloh2005} K. Osterloh, M. Baig, L. Santos, P. Zoller, and M. Lewenstein, Phys. Rev. Lett. \textbf{95}, 010403 (2005).

\bibitem{Ruseckas2005} J. Ruseckas, Juzeliunas, P. \"Ohberg, and M. Fleischhauer, Phys. Rev. Lett. \textbf{95}, 010404 (2005).

\bibitem{Gerbier2009} F. Gerbier and J. Dalibard,  New J. Phys. \textbf{12}, 033007 (2010).

\bibitem{Wang2010} Chunji Wang, Chao Gao, Chao-Ming Jian, and Hui Zhai, Phys. Rev. Lett.  \textbf{105}, 160403 (2010).

\bibitem{Zhu2006} Shi-Liang Zhu, Hao Fu, C.-J. Wu, S.-C. Zhang, and L.-M. Duan, Phys. Rev. Lett. \textbf{97}, 240401 (2006).

\bibitem{Tang2010} E. Tang, J.-W. Mei and X.-G. Wen, arXiv: 1012.2930v2; K. Sun, Z. Gu, H. Katsura, S. Das Sarma, arXiv:1012.5864v1.

\bibitem{footnote1} Note that while the energy spectrum was derived in Ref.\ \cite{aoki:1996}, Fig.\ \ref{fig:butterfly} significantly extends these previous results by showing the \emph{phase diagram} describing the IQHE, which is encoded via the different coloring of the energy gaps.

\bibitem{aoki:1996} H. Aoki, M. Ando, and H. Matsumura, Phys. Rev. B \textbf{54}, 17296(R) (1996).

\bibitem{kohmoto:1989} M. Kohmoto, Phys. Rev. B \textbf{39}, 11943 (1989).

\bibitem{hasegawa:1990} Y. Hasegawa, Y. Hatsugai, M. Kohmoto and G. Montambaux, Phys. Rev. B \textbf{41}, 9174 (1990).

\bibitem{kohmoto:1992} M. Kohmoto, J. Phys. Soc. Jpn \textbf{61}, 2645 (1992).

\bibitem{fukui:2005} T. Fukui, Y. Hatsugai and H. Suzuki, J. Phys. Soc. Jpn \textbf{74}, 1674 (2005).

\bibitem{hatsugai:2006} Y. Hatsugai,  T. Fukui and H. Aoki, Phys. Rev. B \textbf{74}, 205414 (2006).


\bibitem{Lan2011} The relation between $\Delta \sigma_{xy}$, the number of Weyl points and the existence of a flat band has been recently explored by Zhi Hao Lan et al. (in preparation)

\bibitem{min:2006} H. Min \emph{et al.}, Phys. Rev. B \textbf{74}, 165310 (2006).


\bibitem{gerbier:2010} F. Gerbier and J. Dalibard, New J. Phys. \textbf{12}, 033007 (2003).

\bibitem{lin:2009} Y.-J. Lin \emph{et al.}, Nature \textbf{462}, 628 (2009).

\bibitem{mazza1} A. Bermudez, L. Mazza, M. Rizzi, N. Goldman, M. Lewenstein, and M. A. Martin-Delgado, Phys. Rev. Lett. \textbf{105}, 190404 (2010).

\bibitem{mazza2} L. Mazza, M. Rizzi, M. Lewenstein, and J. I. Cirac, Phys. Rev. A \textbf{82}, 043629 (2010).

\bibitem{koehl:2005} M. K\"ohl, H. Moritz, T. St\"oferle, K. G\"unter, and T. Esslinger, Phys. Rev. Lett. \textbf{94}, 080403 (2005).

\bibitem{Umucalilar:2008} R.~O. Umucalilar, H.  Zhai and M.~\"O. Oktel,  Phys. Rev. Lett. \textbf{100}, 070402 (2008).

\bibitem{ho:2000} T.~L. Ho and C.~V. Ciobanu, Phys. Rev. Lett. \textbf{85}, 4648 (2000).







\end{thebibliography}
\end{document}